\documentclass[acmsmall]{acmart}

\newtoggle{show_comments}
\toggletrue{show_comments}
\newtoggle{anonymous}
\toggletrue{anonymous}

\newtoggle{highlight_revisions}
\toggletrue{highlight_revisions}

\newcommand{\changed}[1]{\textcolor{black}
{\iftoggle{show_comments}{#1}{}}}

\newcommand{\finalrev}[1]{\textcolor{black}{\iftoggle{show_comments}{#1}{}}}

\AtBeginDocument{%
  }

\usepackage{colortbl}
\usepackage{xcolor}
\usepackage{csquotes}
\usepackage[most]{tcolorbox}
\tcbuselibrary{listingsutf8}

\setcopyright{cc}
\setcctype{by-sa}
\acmJournal{PACMHCI}
\acmYear{2025} \acmVolume{9} \acmNumber{7} \acmArticle{CSCW467} \acmMonth{11} \acmDOI{10.1145/3757648}

\begin{document}




\title[Regulating Social Media]{Regulating Social Media: Surveying the Impact of Nepali Government's TikTok Ban}

\author{Prerana Khatiwada}
\email{preranak@udel.edu}
\orcid{0009-0008-6965-9504}
\affiliation{%
  \institution{University of Delaware}
  \city{Newark}
  \state{Delaware}
  \country{USA}
}

\author{Alejandro Ciuba}
\affiliation{%
 \institution{University of Pittsburgh}
 \city{Pittsburgh}
  \state{Pennsylvania}
  \country{USA}}
\email{alejandrociuba@pitt.edu}
\orcid{0009-0008-7262-9875}

\author{Aditya Nayak}
\affiliation{%
 \institution{University of Pittsburgh}
 \city{Pittsburgh}
  \state{Pennsylvania}
  \country{USA}}
\email{aditya.nayak@pitt.edu}
\orcid{0009-0009-5517-0932}

\author{Aakash Gautam}
\authornote{Co-senior authors.}
\affiliation{%
 \institution{University of Pittsburgh}
 \city{Pittsburgh}
  \state{Pennsylvania}
  \country{USA}}
\email{aakash@pitt.edu}
\orcid{0000-0001-8023-4648}

\author{Matthew Louis Mauriello}
\authornotemark[1]
\affiliation{%
  \institution{University of Delaware}
  \city{Newark}
  \state{Delaware}
  \country{USA}}
\email{mlm@udel.edu}
\orcid{0000-0001-5359-6520}

\renewcommand{\shortauthors}{Prerana Khatiwada, Alejandro Ciuba, Aditya Nayak, Aakash Gautam, and Matthew Louis Mauriello}

\begin{abstract}
Social media platforms have transformed global communication and interaction, with TikTok emerging as a critical tool for education, connection, and social impact, including in contexts where infrastructural resources are limited. Amid growing political discussions about banning platforms like TikTok, such actions can create significant ripple effects, particularly impacting marginalized communities. We present a study on Nepal, where a TikTok ban was recently imposed and lifted. As a low-resource country in transition where digital communication is rapidly evolving, TikTok enables a space for community engagement and cultural expression. In this context, we conducted an online survey (\textit{N=108}) to explore user values, experiences, and strategies for navigating online spaces post-ban. By examining these transitions, we aim to improve our understanding of how digital technologies, policy responses, and cultural dynamics interact globally and their implications for governance and societal norms. Our results indicate that users express skepticism toward platform bans but often passively accept them without active opposition. Findings suggest the importance of institutionalizing collective governance models that encourage public deliberation, nuanced control, and socially resonant policy decisions.
\looseness -1

\end{abstract}



\received{October 2024}
\received[revised]{April 2025}
\received[accepted]{August 2025}

\maketitle
\section{Introduction}
The rise of social media platforms has revolutionized communication, learning, and business across the globe. Existing research on social media behavior primarily focuses on populations within countries that are economic superpowers and have significant influence over technology infrastructures, such as the U.S. and China, or \finalrev{other} affluent regions in South Asia \cite{wu2023malicious, zannettou2024analyzing, yang2024national}. \finalrev{As a result,} significant gaps remain in understanding how social media platforms are used and adapted in countries with limited resources. This work addresses this gap by presenting perspectives from Nepal, a low-resource country in South Asia. In particular, we examine the effect on people when the Nepali government banned TikTok\footnote{A widely popular video-sharing platform owned by ByteDance, a company based in China.}, allowing us to explore how smaller communities and governments respond to attempts to govern and control technology owned and operated by global mega-corporations.

Despite \changed{the} widespread use and significant impact of \changed{TikTok} in Nepali society, TikTok was banned on November 13, 2023 \cite{bannepal, bannepal2}. The Nepali government attributed the decision to disrupting social harmony and the overall impact on public goodwill \cite{washingtonpost2023tiktok}. 
Further, authorities expressed worries that TikTok, a global platform with extensive data collection practices, might threaten user privacy and national security \cite{nepal2023}. Additionally, there were apprehensions about the platform being used to disseminate misleading information or inappropriate content \changed{\cite{risingnepal2023tiktokban}}. Although official statements mention concerns about data privacy and social harmony, the public was not given a clear and definitive justification for the ban. Considering this, the ban sparked considerable debate among the population and policymakers. This uncertainty has led to various speculations, with some suggesting the ban may stem from geopolitical interests, while others point to internal political dynamics or foreign pressure \cite{adhikari2023navigating}. The ban, \finalrev{however}, was lifted in August 2024 \cite{nbc2023tiktok}.

\changed{Our work builds on this foundation by situating itself within the broader discourse surrounding the TikTok ban in Nepal. } Existing case studies on the TikTok ban primarily examine the Nepali government's motivations behind the ban \cite{adhikari2023navigating, thinkchina2023nepal} and users' reactions, particularly through online comments \cite{lamichhane2024tik}. However, gaps remain in our understanding of users' perspectives about the ban and their experiences of using social media during and after the ban. As the ban was lifted, we do not know about the change in user's attitudes and behavior concerning using a previously-banned platform. 
Our research aims to fill these gaps by 
exploring how Nepali people use and perceive TikTok, particularly in the face of bans and restrictions. We addressed the ambiguity surrounding the exact reasons for the ban and the impacts of the decision, including how people transitioned and adapted. We frame our examination of the topic around the following three research questions.
\looseness -1

\begin{itemize}
    \item  \textbf{RQ1}: \textit{How did people in Nepal adopt TikTok before the ban and for what ends?}
    \item \textbf{RQ2}: \textit{How do they perceive the government's TikTok ban?}   
\item \textbf{RQ3}: \textit{How did they navigate on social media after the ban? }    
    \end{itemize}

To address these research questions, we conducted an online survey with \textit{108} TikTok users in Nepal in August 2024, immediately after the ban was lifted. Our key results show that while TikTok serves as a source of entertainment and a \changed{cultural touchpoint in Nepal}, people in Nepal feel conflicted about the platform due to concerns over content quality and moderation. Although many sought other platforms during the ban, the persistence of dedicated users accessing TikTok despite restrictions shows the platform’s strong influence. 
\looseness -1

The main contribution of our work \changed{are threefold:}
i) we provide empirical insights into how people adapt to and use a system following a ban, ii) we highlight the community behaviors and shifts in interactions that develop when people engage with online platforms (e.g., the emergence of alternative platforms), and \changed{iii) we offer actionable guidance for} policymakers and social media platforms to better understand user behavior and community needs, particularly in collectivist cultures \footnote{A collectivistic perspective highlights the significance of group affiliations, where individuals feel a sense of obligation and interdependence with one another. In this context, identity is shaped by group memberships and relationships, making it essential to understand individuals through their connections within social units \cite{sorensen2009collectivism}.} and resilient communities, through improved governance, moderation, media access policies, and digital literacy efforts.

\section{Related Work}
\label{related_work}

\changed{Before discussing the academic work that informs our study, we briefly situate our study within Nepal's sociopolitical context.} 
\looseness -1

\subsection{\changed{Nepal and Its Global Position}}

\changed{
Nepal is a country characterized by its transitional nature both internally through significant political flux, and externally through pressures to assimilate within the global economic order \cite{sharma2016difficult}. 
It is also a country with low resources, including limited technological infrastructure.
Rampant inequities mar Nepal \cite{nepal2011more}, stirring frustrations with its centralized governance system as it neglects marginalized populations, especially those in remote areas. These issues culminated in a violent ten-year civil war that ended in 2006 \cite{gilligan2011civil}. The movement toppled the monarchy and established a centralized federalist state. However, the civil war did not end the country's political uncertainty, as evidenced by its 14 different prime ministers from 2006 to 2008 \cite{dahal2023strengthening}. The turmoil has delayed Nepal's transition into an industrialized nation, but liberalization produced a gradual shift from a fatalistic worldview towards aspirations of globalization \cite{bista1991fatalism}. This can be seen, for example, in the mass migration of Nepal's population; in 2023, 26.6\% of Nepal's GDP came from foreign remittance \cite{remittance}.
At the same time, Nepal’s sociopolitical landscape remains traditional and guided by state authority, creating tensions around digital governance and morality \cite{shneiderman2016nepal}.}

\changed{This shift into globalization is evident by the growing popularity of international social media platforms like Facebook, Instagram, and TikTok, which have all quickly become integrated into Nepali society. Given Nepal’s lack of robust infrastructure and resources historically, even basic forms of communication remain challenging---especially for Indigenous communities navigating linguistic and technological barriers \cite{CulturalSurvival2023, berkeley2023nepal}. 
Global social media platforms position themselves as accessible and efficient spaces for connection and self-expression, though users still navigate these spaces strategically \cite{mcclain2023life}. 
TikTok, in particular, emerged as a platform for local content creators to share their stories and experiences \cite{ma2021business}, fostering supportive communities and encouraging self-expression \cite {daggett2024digital}. 
It also allowed many Nepali citizens to connect to the global economy by growing brands and businesses via connecting to an international consumer base \cite{krizmatic2023tiktok}.}

\changed{Two key insights from these contextual realities are necessary to situate our findings. First, the central government is the primary policymaker, while provincial and local governments have limited authority, often leading to top-down decisions that may conflict with local priorities and values \cite{dhungana2018public}. This leads to paternalistic governance approaches with frequent frustrations and contestations \cite{byrne2018our}. 
Second, as a country opening up to the global market, Nepal has little leverage to negotiate platform policies with larger corporations like ByteDance, the owner of TikTok. Unlike neighboring India and China, Nepal lacks home-grown alternatives; therefore, its population is used to adapting technologies developed abroad.}
\changed{Keeping this sociopolitical context in mind, we now focus on TikTok, exploring its role as a prominent digital platform, its societal influences, resulting ethical issues, and its international regulatory challenges.} \looseness -1

\subsection{TikTok: Purpose, Application and Scrutiny}
TikTok is a short-form video social media platform owned by ByteDance that has gained worldwide popularity, offering users a space for entertainment, creativity, social, and community interaction \cite{milton2023see, schaadhardt2023laughing, gratification,
schaadhardt2023laughing}. The platform caters to a wide demographic, with a significant portion of its user base aged between 16 and 35 \cite{south}.
Entrepreneurs take advantage of its wide demographic and use it as a public relations platform to promote their businesses via word-of-mouth marketing and to maintain good customer relations \cite{video}.
It is also a well-known tool for forming political and social movements \cite{lee2023introduction, literat2023tiktok}, and has had major impacts on social-political movements such as the \#StopAsianHate movement \changed{\cite{lyu2021understanding}}. All these factors contribute to its high user retention \cite{hedonic} and thus showcase the platform's potential for hosting diverse narratives, affordances, and communities \cite{asianhate}.
\looseness -1

\subsubsection{\changed{Socio-Cultural Variations}}
To comprehend the variations in TikTok usage between different regions, it is important to consider their distinct socio-cultural contexts and user behaviors. A study by \citet{kamran} examined the social media practices of working-class women in Muslim societies and challenges the misconception that women in the Global South use technology in solely utilitarian ways, demonstrating that women in these regions also use TikTok for leisure, self-expression, empowerment, and self-representation. Their results further demonstrate that working-class women find empowerment and a sense of community through TikTok, despite societal stigma, while middle-class women may reject it due to its association with ``low-class'' femininity \cite{kamran}. 
\looseness -1

In Asia, TikTok has been used in diverse ways. 
Public hospitals in China have used TikTok to communicate health-related information to citizens \cite{hospital}. In Indonesia, TikTok is the most popular social media platform, with a significant percentage of the population using it daily for entertainment, communication, and information sharing \cite{sari}. In India, TikTok has influenced cultural practices, like the emergence of lip-sync media \cite{chakravarty2023film}.

TikTok has been evaluated as a tool for English language learning, with many students perceiving it as effective. A study involving semi-structured interviews with ten secondary school students from an urban Malaysian school found that the majority accessed TikTok and felt it enhanced their learning experience by making it more enjoyable \cite{student}. Additionally, TikTok has been instrumental in improving library services and reaching out to users beyond traditional library settings \cite{library}. We also see that TikTok is more popular among individuals with high self-esteem, indicating that in developing countries and the Global South, where self-expression and creativity are esteemed, the platform may attract users looking to exhibit their talents and connect with a global audience \cite{nasidi}. The platform also serves as a space for teenagers to network, collaborate, and negotiate social values, further corroborating the idea that the platform is a hub for self-expression, social activism, and challenging dominant narratives \cite{teenager, le2021s}. 
We build on these previous findings by examining how users in Nepal engage with TikTok, focusing on their usage patterns and content preferences. Beyond the cultural background of our study and the positive effects of the TikTok app, it is crucial to understand the broader (and potentially negative) impacts the platform has had on individuals and society.
\looseness -1

\subsubsection{\changed{Political Issues Surrounding TikTok}}
TikTok has faced scrutiny and legal challenges globally. Concerns have been raised about the platform’s impact on young users, including the spread of mis/disinformation and age-inappropriate content, as well as its potential role in exacerbating mental health issues \cite{cbs, edition}. 
TikTok’s popularity surged during the COVID-19 pandemic, which resulted in further youth user privacy concerns and intensifying debates over the legal framework for children’s data protection \cite{kennedy2020if, marwick2022privacy}.
Additional reports indicate a high prevalence of misleading content on the platform, as discussed in the article \textit{In Case You Haven’t Heard} \cite{heard}. TikTok has also sparked discussion about the responsible portrayal of sensitive historical topics, such as during the \#POVHolocaustChallenge \cite{ebbrecht2022serious}.
The platform’s commercial use by creators and the distribution of online comics for profit has led to legal debates surrounding copyright infringement. Universities have also restricted app access, further reflecting concerns about its user impact and potential security risks \cite{legal}. Given these complexities, examining TikTok’s impact in Nepal is essential, particularly during and after its recent ban, as it was directly prompted by concerns over social harmony and goodwill, content moderation, and data privacy. These reasons reflect broader anxieties about the platform’s influence. Our work contributes to this discussion by understanding how such regulatory actions affect user behavior and cultural practices, especially in a context where these issues have not been thoroughly explored.
\looseness -1

\subsection{Social Media Bans: Impacts and Public Perception}
\label{social-media-bans-impacts-and-public-perception}
Social media bans have significant implications for information dissemination, public discourse, and user adaptation strategies. These restrictions, often enacted in response to security concerns, geopolitical tensions, or policy violations, can shape public opinion and introduce biases in online communication \cite{miller2022dictatorerdogan, surfshark2023}. 
Prior research has shown that news articles reporting on bans and users' reactions on social media can play a significant role in shaping narratives around such bans \cite{borg2021news}. 
Studies on censorship, media bias, and the effects of advertising bans underscore the need to critically evaluate the implications of social media bans on communication, public opinion, and societal norms \cite{golovchenko2022fighting,arias2019does}. This critical lens is particularly relevant in light of recent TikTok bans like India's 2020 ban \cite{kumar2023media}, which ignited global discussions on digital sovereignty and geopolitics. Similar security concerns about TikTok continue to stoke calls for bans in other countries like the U.S \cite{bach2023stitching}, which enacted a short-lasting ban at the beginning of 2025 \changed{\cite{usban}}. In Nepal, the platform raised concerns about social harmony and sparked strong reactions from users \cite{lamichhane2024tik, washingtonpost2023tiktok, bannepal, bannepal2}.

\subsubsection{\changed{Community Impact \& Resistance}}
\changed{A key dimension to social media bans is their effect on user communities. While much research has focused on the legal frameworks, policy enforcement, and political justifications for said bans (e.g., \cite{golovchenko2022fighting, McAlister_Beatty_Smith-Caswell_Yourell_Huberty_2024, Zhuravskaya_Petrova_Enikolopov_2020, Zeng_Raess_Musgrave_2025}), fewer studies examine how users navigate and resist such restrictions. Studies on circumvention practices, such as the use of VPNs and alternative platforms, demonstrate how social media bans often lead to adaptive behaviors rather than absolute compliance \cite{gebhart2017internet}.}
\changed{Similarly, research on mass collective action by online community leaders shows how users strategically mobilize and leverage platform dependencies to push back against restrictive policies. This was evident in the 2015 Reddit moderator protest, where moderators of 2,278 subreddits collectively disabled their communities to pressure Reddit into negotiating over their demands
\cite{ matias2016going}.}
\changed{Building on this, a recent systematic review by \citet{jiang} highlights how content moderation—especially during bans—involves tradeoffs between values, moderation styles, and stakeholder needs, which shape how users respond, resist, or adapt to platform constraints.}
\changed{Given the drastic effects bans have on communities and the resulting friction}, it is important to understand how bans affect broader ideals that are vital for informed decision-making and democracy, such as information flow, public discourse, and other societal norms.
\changed{Compounding these issues, 
Ackerman's socio-technical gap \footnote{The socio-technical gap, introduced by Ackerman (2000) \cite{ackerman2000intellectual}, refers to the disconnect between human requirements (socio-requirements) and technical solutions in CSCW. He emphasizes that human activities are flexible and contextual, while computational mechanisms are rigid. Ackerman calls for HCI and CSCW to recognize and address this gap as an intellectual mission.} framework highlights the disconnect between technical controls and complex human behaviors \cite{ackerman2000intellectual}. This is relevant in understanding digital restrictions for developing regions. Applying this lens to Nepal's TikTok ban, our work analyzes how platform bans impact user migration patterns and digital resilience strategies.}
\looseness -1

\subsubsection{\changed{Migration and User Adaptation}}
\label{migration-and-user-adaption}
The migration of individuals between social media platforms is a dynamic process driven by various factors. 
Users often engage with multiple platforms instead of relying on a single one due to overlapping media functionalities
\cite{polymedia},
while distinct platform features further encourage switching based on evolving needs and preferences \cite{chinese}. 
Factors such as digital architecture changes, platform reprogramming, and user experience alterations can prompt individuals to seek alternative platforms that better align with their requirements \cite{bossetta2018digital, bossetta2020scandalous}.
\changed{Research on Twitter to Mastodon migration examines user behavior following platform ownership changes, finding that while users migrate quickly, their retention depends on behavioral adaptation and platform architecture differences \cite{jeong2024exploring}.} Motivations such as social interaction, entertainment, information sharing, status achievement, and time utilization have been identified as key drivers for using social media platforms \cite{threat}. These factors compound with additional affordances such as content diversity, interactivity, and the availability of various information sources to influence users' decisions to switch platforms \cite{typology}. Users perceive significant benefits from transitioning between platforms, including access to varied information, content choice, exposure to diverse perspectives, and enhanced interactivity \cite{typology}. 
\looseness -1

\changed{Beyond user-driven platform switching, forced migration due to platform bans presents unique challenges and adaptation strategies. \citet{fiesler2020moving} highlights how fandom communities relocate across platforms when technical or policy changes disrupt their activities. Similarly, \citet{horta2021platform} examine how toxic communities, such as r/The\_Donald \footnote{\changed{
r/The\_Donald: A pro-Trump subreddit known for political trolling, conspiracy theories, and extremist rhetoric; associated with the alt-right movement and banned by Reddit in 2020 for repeated policy violations.}} and r/Incels \footnote{\changed{r/Incels: A subreddit for "involuntary celibates" built around misogynistic and fatalistic ideologies like the "black pill"; banned by Reddit in 2017 due to its links to hate speech and real-world violence.}}, responded to Reddit bans, finding that while bans effectively reduce activity, they can also promote ideological radicalization. These findings suggest that while migration can be a survival strategy for communities, it also introduces new risks and transformations. Platform bans can lead to infrastructural migration, where users do not simply shift to an alternative platform but rather reassemble their digital presence across multiple services. \citet{zhang2022chinese} illustrates this concept through the WeChat ban, which forced users to fragment their communication channels rather than relocate entirely. This concept is particularly relevant in restrictive digital environments, where platform bans disrupt social, economic, and political interactions.} 
\looseness -1

\changed{Building on these insights, our research examines Nepal’s TikTok ban from the perspectives of people in a developing country where externally developed social media platforms are increasingly shaping everyday life. Unlike prior work that largely focused on Western contexts, we explore how Nepali users adapt through alternative platforms and shift their content-sharing behaviors. 
}
\changed{
As our previous discussion showed,
platform restrictions, often driven by government regulation and political agendas, shape access and frequently prompt user migration, whether framed as digital sovereignty, content control, or ideological enforcement. This necessitates a closer look at the role of political power in media regulation, as an examination of a governmental ban is incomplete without a discussion of its surrounding politics.}

\subsection{The Influence of Politics and Power on Media Platforms}
\changed{Social media is not just a neutral communication space; rather, it is deeply embedded in power dynamics \cite{massidda20203, 
khosravinik2016critical}.} The relationship between politics and media power is crucial as social media shapes public discourse and political engagement \cite{chadwick2015politics, nahon2015there}. Governments and political entities use these platforms for censorship, and to influence/control public perception \cite{stieglitz2013social, kavanaugh2011social}.
\changed{Across the world, social media has become a battleground for political narratives, where platform governance becomes intertwined with state power, propaganda, and public opinion manipulation \cite{wang2019china,saaida2023role}.} For example, strict censorship policies in China ensure that only state-approved messages dominate online spaces, controlling the information that citizens can access \cite{king2013censorship,king2017chinese}. Similarly, during the 2016 U.S. election, targeted ads and fake news on Facebook influenced public perception and voter behavior \cite{liberini2016politics}. 
\looseness -1

Algorithms play a key role in this dynamic, as they can amplify certain voices while suppressing others, often reinforcing existing societal inequalities \cite{zimmer2019echo}. Considering low-resource settings, where people may not have access to various news sources, social media becomes a primary way to get information \cite{ali2011power}. As seen in places like Myanmar, social media and their algorithms were used both as tools for misinformation and protest organization during the 2021 military coup, contributing to public confusion and unrest \cite{brooten2020media}. \changed{Examining platform bans and migration from Section \ref{social-media-bans-impacts-and-public-perception}, we see these often stem from governmental interventions that shape digital spaces to align with political, security, or ideological objectives. Thus, the TikTok ban in Nepal, much like those in India \cite{kumar2023media} and the U.S. \cite{clausius2022banning}, raises questions about how governments justify such restrictions, whether for national security, misinformation control, or political suppression.} Our work contributes to this discussion by examining how power dynamics in media shape digital engagement \changed{and the broader implications of government intervention on public discourse, societal norms, and digital rights.}
\looseness -1

\section{Methodology}
\label{methods}
We explain the design and preparation of the study materials, including the survey structure and methodology, and then outline the research process, recruitment, and steps to ensure validity and cultural relevance. The study was approved by the University of Delaware’s Institutional Review Board (IRB) under protocol number 2200126-1.

\subsection{Study Materials}
\label{sec:materials}
\changed{Our study was conducted through an online survey.} All study materials, including consent forms and surveys, were initially prepared in English. Then, after multiple iterations and expert reviews, the English-to-Nepali translation was carried out using Google Translate \footnote{\url{https://translate.google.com/}} and a Unicode Nepali converter \footnote{\url{https://www.ashesh.com.np/nepali-unicode.php}}. Google Translate sometimes gave literal translations, so we adjusted them based on discussions between two native Nepali speakers on our research team to ensure appropriate translations that were in line with everyday spoken Nepali rather than literal translations. 
In the second phase, we worked with Prakash Bhattarai from Nepal’s Centre for Social Change \footnote{\url{https://socialchange.org.np/}} (CSC---a non-profit research and advocacy institute) to carry out a cultural sensitivity analysis
 \footnote{The phrase “cultural sensitivity” is multi-ordinal, meaning it can have different interpretations and implications depending on the considered level of abstraction. Rychlak \cite{rychlak1968philosophy} noted that many psychological terms exhibit this characteristic, where their meanings shift based on context or its discussed aspects.}. This ensured that our survey was respectful and appropriate for the target population. We chose this organization for its expertise in Nepalese socio-political dynamics, including conflict transformation, democracy, governance, migration, labor and employment, civic space development, and public policy. Overall, the translation process involved direct linguistic conversion and cultural adaptation of terms to resonate with the Nepali audience, as certain expressions may have different connotations in Nepal. To manage participant compensation, we partnered with a local contact in Nepal, as paying participants internationally proved to be \changed{logistically challenging}. The local contact was onboarded as a foreign vendor through the \changed{university’s procurement process, which involved registration, credentials verification, and ensuring compliance with institutional guidelines}.
\changed{All of our study materials are in Supplemental Note 3.}
\looseness -1

\subsubsection{Survey design, structure and components}
The survey was developed in Qualtrics by two team members, and reviewed by a third who, while not of Nepali nationality, reviewed the English version for clarity and consistency before it was translated to Nepali. 
\changed{Survey questions were informed by an initial review of policy documents and advocacy materials from the governing body and key data organization, including 
Body and Data \footnote{\url{https://bodyanddata.org/}}, a digital rights organization in Nepal, and the
Centre for Social Change, as well as public discourse in opinion pieces, blog posts, commentary and local news (e.g, Kathmandu Post, Online Khabar). 
The ``Status of Digital Rights in Nepal: A Review and the Media Monitoring Report'' \cite{bodyanddata2024mediamonitoring} by Body and Data was also crucial in informing our framework as it highlights key digital rights issues, including online privacy and data protection, online freedom of expression, surveillance, and censorship. To further contextualize the research and understand how the issue is perceived at the community level, we held an informal Zoom session with Prakash Bhattarai of the Centre for Social Change. Insights from this session, along with findings from a related study on public responses to the TikTok ban in Nepal \cite{lamichhane2024tik} informed us when developing survey questions, making them grounded in both local realities and current discourse. Specifically, we drew from the themes and categories identified in Lamichhane’s qualitative analysis of YouTube comments, such as emotional responses, concerns over livelihood, and expressions of digital rights, to shape the content and tone of our questions \cite{lamichhane2024tik}.
We also incorporated insights from expert commentary and policy analysis, such as \citet{jha2024tiktok}’s examination of the political, social, and regulatory context behind Nepal’s TikTok ban, which highlighted issues like social harmony, cybercrime concerns, business disruption, and civil liberties. 
}

\begin{figure}[t]
    \centering
    \includegraphics[width=\textwidth]{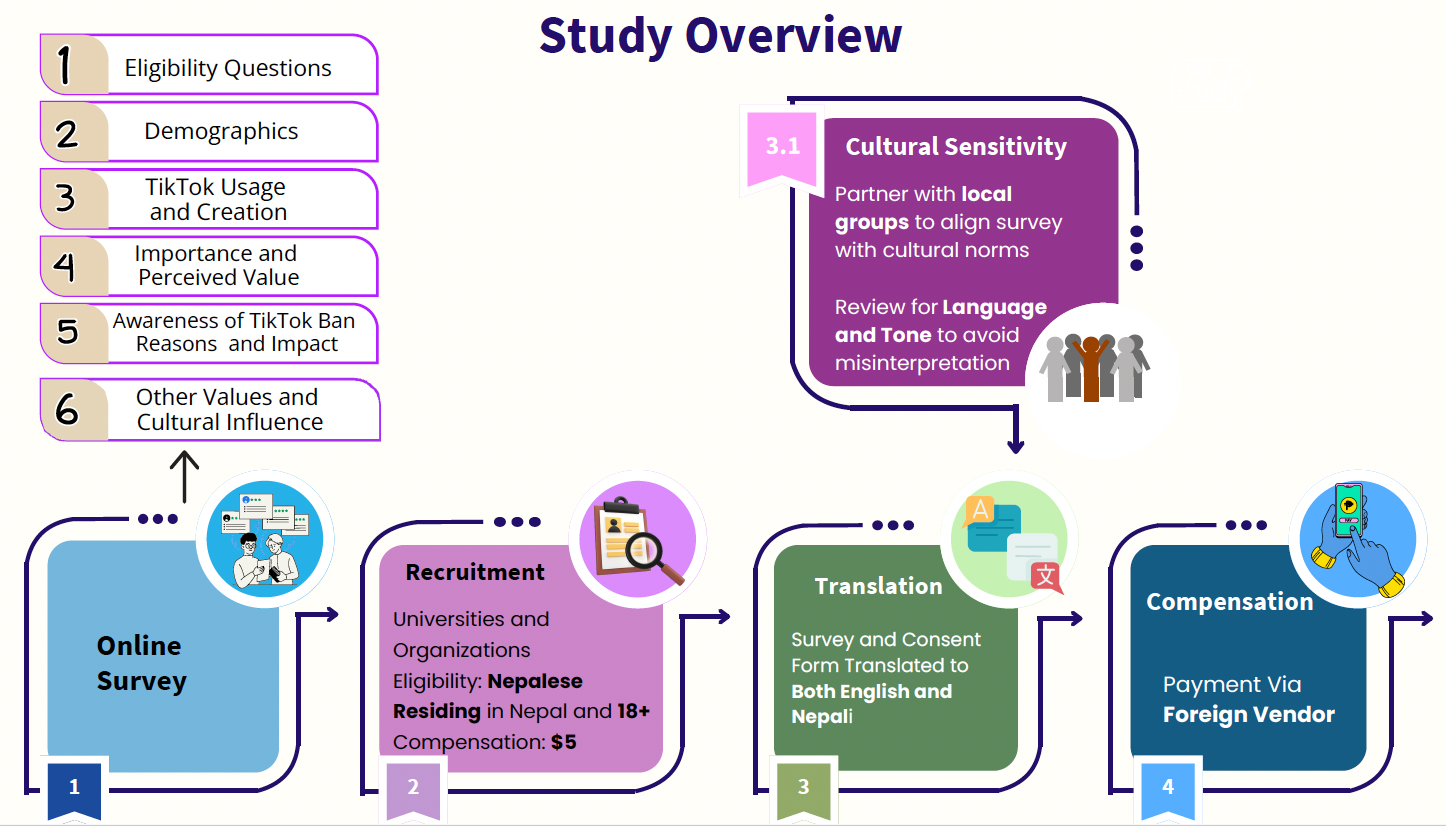} 
    \caption{\finalrev{Overview of Study Workflow}}
    \label{fig:image}
  \vspace{-1.2em} 
\end{figure}

We used a mix of closed-ended and open-ended questions to collect both quantitative and qualitative data.
Questions such as frequency of TikTok use, purpose of use (e.g., entertainment, business, socializing), and perceptions of TikTok’s impact (on personal life, business, and social relationships) were explored through multiple-choice and Likert-scale questions. As a part of the Likert scale survey, participants were asked to rate their agreement with statements like “The ban on TikTok was a violation of freedom of expression” and “How satisfied were you with the reasons provided for banning Tiktok?” We also designed open-ended questions to capture nuanced personal experiences and feelings, such as questions like “Do you think alternative apps are equivalent or comparable to TikTok? Please explain your reasoning?” Demographic questions were also included to understand participants’ background (e.g., age, gender, occupation, education level, employment,  internet usage (Wi-Fi or mobile data), and place of residence), which helped us analyze trends across different user groups.

Our reviewed literature informed a five-section online survey: 1) TikTok \finalrev{U}sage and \finalrev{C}reation Habits, 2) Importance and \finalrev{P}erceived \finalrev{V}alues, 3) Perceptions Towards the \finalrev{B}an, 4) Social Media Transition and Usage Habits, and 5) Demographics. \changed{The survey questionnaire, in both Nepali and English, is provided in Supplemental Note 3.3.}

\subsubsection{Cultural Sensitivity Analysis and Language Selection}
As mentioned in the Section \ref{sec:materials}, a cultural sensitivity analysis was conducted to ensure the language used was appropriate and not offensive to Nepali participants. For example, in one question that initially stated, \textit{``TikTok was crucial for my small businesses to reach a wider audience and grow,''} we reconsidered the phrasing. In Nepal, calling a business ``small'' may come across as dismissive or diminishing, whereas in the U.S., ``small business'' often carries a positive connotation. We removed the qualifier ``small'' and kept the term ``business'' singular to better reflect how Nepali people perceive their ventures. 
Participants select their preferred language at the start of the survey. Participants selecting English saw only the English questions, while those choosing Nepali received questions in Nepali with English translations to support clear understanding.
\looseness -1

\subsubsection{Improving Survey Quality}
We incorporated two attention-check questions throughout the survey to maintain participant engagement and to filter out respondents who failed to answer them correctly. One question asked participants, \textit{Imagine you are at a picnic with friends. You have brought along a basket filled with different fruits. If you’re paying attention to this survey, please select “I can name three fruits” from the options below.} We also applied 50-character limit to open-ended questions, so that responses were thoughtful and of good quality. We conducted a pilot test of the survey with three people (2 Nepali and 1 Non-Nepali) who were not part of the research team.
By getting feedback from outside our team, we were able to spot any confusing or biased wording that could have affected how people answered. One pilot participant raised concerns about the question regarding uploading files, recommending that it be removed or made optional. Given TikTok’s new bookmark feature, they suggested rephrasing the question to “Have you saved or bookmarked the content?” Another added, \textit{“Some questions seemed to assume that everyone uses TikTok in the same way. They didn’t account for how different age groups might use the platform differently.”} We designed the survey using conditional logic to display questions based on participants’ earlier responses. After pilot feedback, we revised questions for clarity, using simple language and avoiding technical terms.
\looseness -1

\subsection{Recruitment}
We conducted an online survey in Fall 2024, recruiting \textit{108} participants based on a power analysis ran in G*Power \footnote{\finalrev{G*Power is a free, open-source statistical power analysis tool,
widely used in social and behavioral sciences to calculate required sample sizes for various statistical tests.
}} \cite{kang2021sample}.
\changed{Our a priori power analysis used (two-tailed, $\alpha$ = 0.05, power = 0.80) with an expected effect size of \textit{r = 0.26}, resulting in a required sample size of 113 participants. Our actual sample of 108 participants closely meets this threshold to have adequate power for the planned correlation analysis.}
\changed{We collected data over three weeks during the key period after the TikTok ban was lifted,}
capturing participants’ immediate reactions, behavior changes and coping strategies as they re-engaged with the platform.
\looseness -1

Participants were recruited via word of mouth, social media, and outreach at four local colleges, using contacts from our research team. We combined targeted outreach with snowball sampling, encouraging participants to share the survey within their networks. However, the sample was primarily one of convenience. The recruitment posts and ads outlined the purpose of the study and a link to the survey questionnaire.
\changed{Over the course of a month,} we compensated participants with \$5 for completing the survey, \changed{based on the exchange rate at the time of payment,} which we consider appropriate given the cost of living in Nepal. This payment, though modest, aligned with regional norms and acknowledged the participants’ time and contributions. 
Our study included self-identified regular TikTok users who were at least 18 years old, of Nepali nationality, and residing in Nepal \changed{at the time of the study.} See Figure \ref{fig:image} for study details.

\subsection{Participants}
Participant demographics show a mostly young population, with 53.7\% aged between 18-24 years. A slight majority (54.63\%) identified as male, while 44.44\% identified as female, with only one participant (0.93\%) preferring to self-describe. 
Geographically, most participants (77.78\%) resided in urban areas, while very few (5.56\%) lived in rural areas. 
\changed{Our household income data shows a significant proportion of respondents preferred not to disclose their earnings, being the largest category. Among those who did, most fall within the lower to middle-income brackets, with monthly income between NPR 20,000 - NPR 40,000 ($\approx$USD \$145 - \$290)  being the most common, followed by NPR 40,000 - NPR 60,000 ($\approx$USD \$290 - \$435). Fewer respondents reported monthly income below NPR 20,000 (USD \$145) or above NPR 100,000 (USD \$725).} 
 Participant demographics are presented in Table \ref{tab:dem}.
 \looseness -1

\begin{table}[t]
\centering
\caption{Self-Reported \finalrev{Demographic Breakdown of the Study Sample}
}
\small
\begin{minipage}{0.3\textwidth}
\centering
\begin{tabular}{|l|l|}
\hline
\rowcolor{gray!30} \textbf{Age} & \textbf{N (\%)} \\
\hline
\rowcolor{gray!10} 18-24 & 58 (53.7\%) \\
\rowcolor{white} 25-34 & 49 (45.37\%) \\
\rowcolor{gray!10} 54+ & 1 (0.93\%) \\
\hline
\end{tabular}
\end{minipage}%
\hfill
\begin{minipage}{0.3\textwidth}
\centering
\begin{tabular}{|l|l|}
\hline
\rowcolor{gray!30} \textbf{Gender} & \textbf{N (\%)} \\
\hline
\rowcolor{gray!10} Male & 59 (54.63\%) \\
\rowcolor{white} Female & 48 (44.44\%) \\
\rowcolor{gray!10} Self-describe & 1 (0.93\%) \\
\hline
\end{tabular}
\end{minipage}%
\hfill
\begin{minipage}{0.3\textwidth}
\centering
\begin{tabular}{|l|l|}
\hline
\rowcolor{gray!30} \textbf{Occupation} & \textbf{N (\%)} \\
\hline
\rowcolor{gray!10} Student & 57 (52.78\%) \\
\rowcolor{white} Full-time & 34 (31.48\%) \\
\rowcolor{gray!10} Part-time & 8 (7.41\%) \\
\rowcolor{white} Unemployed & 4 (3.70\%) \\
\rowcolor{gray!10} Self-employed & 3 (2.78\%) \\
\rowcolor{white} Homemaker & 2 (1.85\%) \\
\hline
\end{tabular}
\end{minipage}%

\begin{minipage}{0.3\textwidth}
\centering
\begin{tabular}{|l|l|}
\hline
\rowcolor{gray!30} \textbf{Education} & \textbf{N (\%)} \\
\hline
\rowcolor{gray!10} Bachelor's & 59 (54.63\%) \\
\rowcolor{white} High school & 35 (32.41\%) \\
\rowcolor{gray!10} Master's & 13 (12.04\%) \\
\rowcolor{white} No education & 1 (0.93\%) \\
\hline
\end{tabular}
\end{minipage}%
\hfill
\begin{minipage}{0.3\textwidth}
\centering
\begin{tabular}{|l|l|}
\hline
\rowcolor{gray!30} \textbf{Marital Status} & \textbf{N (\%)} \\
\hline
\rowcolor{gray!10} Single & 73 (67.59\%) \\
\rowcolor{white} Married & 17 (15.74\%) \\
\rowcolor{gray!10} Relationship & 18 (16.67\%) \\
\hline
\end{tabular}
\end{minipage}%
\hfill
\begin{minipage}{0.3\textwidth}
\centering
\begin{tabular}{|l|l|}
\hline
\rowcolor{gray!30} \textbf{Residence} & \textbf{N (\%)} \\
\hline
\rowcolor{gray!10} Urban & 84 (77.78\%) \\
\rowcolor{white} Suburban & 18 (16.67\%) \\
\rowcolor{gray!10} Rural & 6 (5.56\%) \\
\hline
\end{tabular}
\end{minipage}%
\vspace{10pt}
 
\label{tab:dem}
\end{table}

\subsection{Overview of Analysis Method}
\changed{Our online survey received a total of 251 responses, with 108 retained valid responses\footnote{\changed{Any exclusions were due to ineligibility or incomplete data}.}.}
There were no instances of speeding or straight-lining \changed{\cite{zhang2014speeding}}, and the survey maintained a good overall completion rate. After filtering out low-quality responses, our final dataset included 84 valid responses from participants who selected English and 24 from those who chose Nepali. We then coded, anonymized, and stored the data from the demographics and survey responses using participant IDs.
\looseness -1

We analyzed the quantitative survey items using descriptive statistics (i.e., mean, median, frequencies, and standard deviation). 
\changed{Based on the nature and distribution of the data, we used chi-square tests and correlation analysis to examine variable relationships and significant patterns.}
\changed{In the case when} participant uploaded files in response to questions like ``Do you have a favorite TikTok video saved or bookmarked?'' and ``Do you have photos or videos of the concerning posts/content? Please upload them here. The submissions were downloaded for further analysis.
\looseness -1

We thematically clustered the responses to the free-text questions \cite{braun2012thematic}.
\changed{Given the exploratory nature of our study, the thematic analysis focused on identifying emergent patterns over testing predefined coding schemes.} Consistent with qualitative approaches focusing on interpretive depth, we did not calculate inter-rater reliability \cite{mcdonald2019reliability}. Instead, the coding process first focused on arriving at a collaborative agreement among the coders on a sample of codes. Two authors coded 60 responses together to reach a consensus on the coding scheme. Following this, the overarching themes were developed via inductive coding \cite{thomas2003general}, allowing themes and codes to emerge naturally from the participant responses. 
\changed{Through this iterative process, six major themes emerged (see Table \ref{tab:thematic-analysis}.)}
\changed{As an example, merging through low-level codes, we arrived at subcategories such as ``TikTok is addicting'' and ``TikTok is vulgar'', which we then grouped under the  broader ``Entertainment'' subcategory due to their focus on content-related experiences. 
Subcategories like ``educational content'' and ``misinformation'' were merged into the ``Information'' theme \footnote{See \changed{Supplemental Note 1 for more subtheme examples.}}.
}
\changed{In cases of disagreement, the two coders convened to compare interpretations, discuss discrepancies, and refine the codebook to ensure consistency and clarity in the final themes.} Thematic clusters derived from the codes form the basis of our qualitative results \changed{presented in Section \ref{altplat}}. 
\looseness -1

\section{Findings}
\label{Results}
\begin{figure}[t]
    \centering
\includegraphics[width=\textwidth]{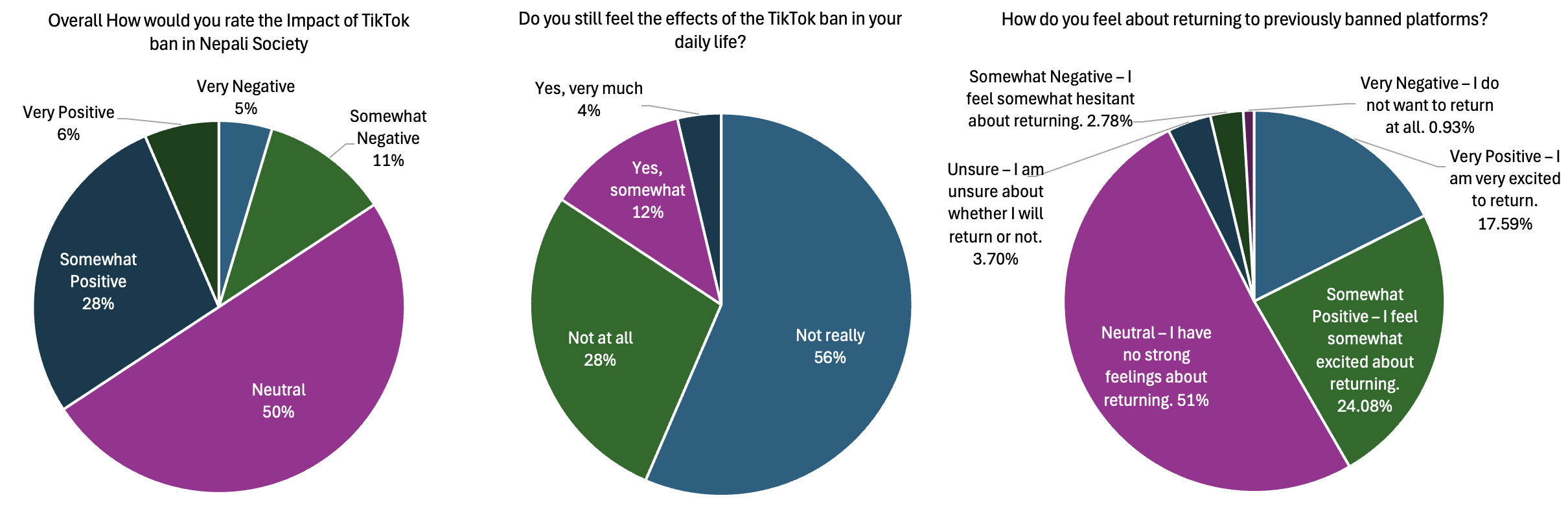}
    \caption{\centering Overall Nepali Perceptions on Social Media Bans: Key Themes and Societal Impact}
    \label{teaserfigure}
\end{figure}

We present findings from our survey on TikTok usage patterns,
user motivations and perceptions of the ban. 
Table \ref{tab:usage} shows TikTok had a strong user base in Nepal with frequent pre-ban use and a notable role in daily routines. The key overarching findings produced by this work are illustrated in Figure~\ref{teaserfigure}. \changed{We begin by examining the perceived benefits and societal impact of TikTok in Nepal.}

\begin{table}[b]
    \centering
    \small
    \caption{TikTok Usage Behavior and Duration Before the Ban}
    \label{tab:tiktok_usage}
    \begin{tabular}{|l|c|c|}
        \hline
        \textbf{Category} & \textbf{Frequency} & \textbf{\%} \\ 
        \hline
        \multicolumn{3}{|c|}{\textbf{Content Engagement}} \\ 
        \hline
        Creating and sharing content & 7 & 6.48\% \\ 
        Both creating and watching content & 45 & 41.67\% \\ 
        Watching content only & 56 & 51.85\% \\ 
        \hline
        \multicolumn{3}{|c|}{\textbf{Duration of TikTok Usage}} \\ 
        \hline
        More than 5 years & 9 & 8.26\% \\ 
        3-6 months & 15 & 13.76\% \\ 
        6-12 months & 18 & 16.51\% \\ 
        3-5 years & 18 & 16.51\% \\ 
        1-3 years & 48 & 44.04\% \\ 
        \hline
        \multicolumn{3}{|c|}{\textbf{Frequency of Usage Before Ban}} \\ 
        \hline
        Rarely & 7 & 6.42\% \\ 
        Once a day & 12 & 11.01\% \\ 
        A few times a week & 14 & 12.84\% \\ 
        Multiple times per day & 75 & 68.81\% \\ 
        \hline
        \multicolumn{3}{|c|}{\textbf{Daily Time Spent Before Ban}} \\ 
        \hline
        More than 3 hours & 13 & 11.93\% \\ 
        2-3 hours & 17 & 15.6\% \\ 
        15-30 minutes & 22 & 20.18\% \\ 
        30-60 minutes & 23 & 21.1\% \\ 
        1-2 hours & 33 & 30.28\% \\ 
        \hline
    \end{tabular}
    \label{tab:usage}
\end{table}


\subsection{Perceived Impact of TikTok Usage}
\label{sec:perceivedimpacts}
Participants rated both positive and negative aspects of TikTok using a scale from 1 (strongly disagree) to 7 (strongly agree).
Results indicate that TikTok was primarily valued for entertainment (M = 5.65); other areas received more moderate or lower ratings (see Table \ref{tab:positive}). \changed{Thus, most participants were engaging with TikTok more for passive consumption than active creation; a pattern consistent with the fact that the majority of respondents did not identify as content creators (see Table \ref{tab:usage})}. 

However, concerns were also strong (see Table \ref{tab:negative}). TikTok was seen as distracting youth from responsibilities (M = 5.50), exposing minors to inappropriate content (M = 5.45), and contributing to misinformation (M = 4.83), cyberbullying (M = 4.87), and promoting other harmful societal behaviors. These findings suggest, 
despite the platform’s global appeal, in societies like those in Nepal, where cultural values and social responsibilities are closely intertwined \cite{relaxgetaways2024}, the platform exacerbates existing social challenges.

\begin{table}[t]
    \centering
    \small
    \caption{Mean and Standard Deviation of Perceived Positive Impacts of TikTok in Nepal}
    \begin{tabular}{@{}lcc@{}}
        \toprule
        \textbf{Statements:
        To what extent do you agree with the following statements? } & \textbf{Mean ± Std.} \\ \midrule
        1. TikTok provided me with entertainment & 5.65 ± 1.35 \\ 
        2. TikTok promoted my creativity and self-expression & 4.71 ± 1.87 \\ 
        3. TikTok facilitated my community engagement and social interaction & 4.57 ± 1.90 \\ 
        4. TikTok offered me educational or informative content & 5.20 ± 1.51 \\ 
        5. TikTok inspired me to become more aware of social issues and  encouraged social change & 5.13 ± 1.55 \\ 
        6. TikTok fostered a sense of belonging or connection for me & 4.13 ± 1.79 \\ 
       7. TikTok supported my mental health and well-being & 4.01 ± 1.99 \\ 
        8. TikTok played a vital role in preserving and promoting my local  cultures and traditions & 4.68 ± 1.71 \\ 
        9. TikTok was crucial for my business to reach a wider audience and grow & 4.77 ± 1.79 \\ 
        \bottomrule
    \end{tabular}
    \label{tab:positive}
\end{table}

\begin{table}[b]
    \centering
    \small
    \caption{Mean and Standard Deviation of Perceived Negative Impacts of TikTok in Nepal}
    \begin{tabular}{@{}lcc@{}}
        \toprule
        \textbf{Statements: To what extent do you agree with the following statements?} & \textbf{Mean ± Std.} \\ \midrule
        1. TikTok promoted values and trends that did not resonate with Nepali culture & 4.52 ± 1.59 \\ 
        2. TikTok promoted immoral values & 4.60 ± 1.47 \\ 
        3. TikTok distracted young people from their studies or responsibilities & 5.50 ± 1.48 \\ 
        4. TikTok has contributed to the spread of misinformation in Nepal & 4.83 ± 1.65 \\ 
        5. TikTok exposes young users and kids to inappropriate content unsuitable for their age & 5.45 ± 1.33 \\ 
        6. TikTok fosters hate speech, cyberbullying, negatively impacting users' mental well-being & 4.87 ± 1.53 \\ 
        \bottomrule
    \end{tabular}
    \label{tab:negative}
\end{table}

\subsection{Motivations and Actions Surrounding the Ban}
\label{sec:useafterban}


\changed{Building on the perceived impacts, we now explore how user motivations and experiences shaped their behavior in response to the TikTok ban. With respect to usage,} 
36.36\% stopped using the app after the ban, while 13.64\% had already stopped beforehand. Despite the restriction, 12.12\% continued using it, and nearly a quarter, 23.48\% returned once the ban was lifted—including 26\% of those who had quit even before the ban. After the ban was lifted, 17.65\% of returning users said their reason for using TikTok had changed, often shifting to new purposes such as watching brand promotions. Usage patterns also shifted, with a noticeable reduction in daily time spent, as shorter usage durations (15-30 and 30-60 minutes) became more frequent, and fewer users reported spending over 2 hours per day, suggesting more conservative habits among returning users. 
\looseness -1


\changed{To complement the quantitative trends discussed above, here we outline participants' reasons for quitting TikTok before the ban, using it during the ban, and returning afterward.}

\subsubsection{Reasons for Stopping TikTok Usage}
Participants expressed concerns about time management and a lack of meaningful or productive tasks on TikTok as major reasons for deciding to quit the platform before the ban. For example, a participant shared: 
\begin{displayquote}
\textit{“I wasn’t creating anything—just endlessly consuming content. I found myself glued to the screen and putting off my responsibilities. At first, it was entertaining, but eventually, I felt drained and realized the videos offered little value.”}
\end{displayquote}

Many participants also felt disappointed by repetitive, low-quality content and expressed concerns that short-form videos hindered their attention span and ability to focus on meaningful tasks, \textit{“I didn’t find it beneficial in the context of Nepal. Most of the people are into watching TikToks that are entertaining. I wanted to create something related to educational...''} 



Some expressed concerns about inappropriate content on TikTok. A participant shared, 
\textit{``People spread cringe and nudity nowadays.''} Another participant said, \textit{``It is addictive. And I use reels. The content from TikTok gets filtered to reels anyway. That's why I stopped.''}
In all of these, we note participants' critical reflection on the interactions TikTok afforded and their agency to stop using TikTok when they found that it did not align with their values. 

\subsubsection{Reasons for Continued Engagement Despite Ban}


Entertainment was one of the most salient reasons participants accessed TikTok post-ban, as a participant shared \textit{``Just when I am free, I check out TikTok rarely to enjoy the content.”}
They also saw potential for educational and business-related benefits by being on TikTok, which made one of the respondents question the ban:
\begin{displayquote}
\textit{“Watching creative content on TikTok was entertaining and informative. I believe banning it was misguided, as it could have [been] targeted [on] individuals creating vulgar content instead. TikTok is a valuable platform for sharing ideas and has significantly helped small businesses thrive.”} 
\end{displayquote}
Others opposed the ban and continued using TikTok in protest. One participant shared, \textit{“I felt banning TikTok was a politically motivated action, and since I still enjoyed the content, I kept using it”.}

\subsubsection{\changed{Circumventing the TikTok Ban}}
\label{sec:vpn}
\changed{Having explored the motivations behind continued TikTok use despite the ban, we now turn to a closer examination of how many users actually persisted in using the platform}. Fifteen participants (13.88\%), reported continuing to use TikTok despite the ban. \changed{This subset employed a range of technical strategies} to bypass the restriction: Six (6) reported using VPNs, another six (6) opted for changing their DNS settings, often using Google DNS or Cloudflare DNS to bypass the ban. The others (3) sideloaded the app directly or accessed the web version. Participants reported learning these workarounds to come back to TikTok:
\begin{displayquote}

``\textit{. . . After the ban, I switched to Instagram for my fix of reels, but I missed the fun and creativity of TikTok. So, I found a way around the ban and continued to enjoy, even if I don't spend as much time there anymore.}''

\end{displayquote}
\changed{Notably, those who bypassed the ban were from urban areas, had reliable home WiFi, and had either bachelor's or master's degree. Most were either students or employed, and several reported relatively high monthly incomes (ranging from NPR 40,000 to 120,000). This suggests a class-based dynamic in access to both the technological know-how and infrastructure needed to circumvent government regulation.}
\changed{We could also hear this in a respondent's concern about how the TikTok ban disproportionately affected women,}
\begin{displayquote}
``\textit{\changed{TikTok was a very welcoming app for women creators in the sense that there were a lot of women creators in the appp [app], which I think has been reduced when transitioning from TikTok. The transition has cut a lot of small creators like small business owners, less digital-savvy users, and women as well.}''}
\end{displayquote}
\looseness -1

\subsubsection{Resuming Engagement Post-Ban}
\label{engagementpostban}
The 23.48\% who resumed TikTok after the ban waited for official approval and shared that they were motivated to return in order to learn about what is trendy, \textit{``I was eager to explore the trends and viral content I missed.''}
Catching up with the content was seen as particularly important. One participant, who did not even uninstall the app during the ban, shared, 
\begin{displayquote}
\textit{``I had never uninstalled it, so I wanted to catch up on all the TikTok content I lost track of when it was banned.''} 
\end{displayquote}

In some cases, the lifting of the ban seems to have drawn attention to the value of TikTok in being part of the trending cultural practices, bringing in new users:
\begin{displayquote}

\textit{``I started using TikTok after the government lifted the ban, and my reasons were influenced by a mix of curiosity, social pressure, and a desire to connect with a wider cultural moment. When the platform became accessible again, I couldn’t help but feel intrigued by all the buzz surrounding it. Throughout the ban, I had heard about TikTok’s rise in popularity. Friends and social media platforms often discussed the trends, viral challenges, and creative content that TikTok was known for, which made me wonder what I had been missing. As soon as the app was available to me again, I decided to dive in and explore it firsthand.''}
\end{displayquote}

A few also highlighted TikTok's role in maintaining connections with friends and families, especially those living overseas, as one said, \textit{``...You know, TikTok has been such an important way for me to connect with my friends and family, especially those living overseas. Umm, it really keeps those connections alive.''}
This is particularly pertinent to the Nepali context where mass migration is the norm and telecommunication infrastructure beyond larger platforms like TikTok remains limited.
\looseness -1



\begin{figure}
    \includegraphics[width=\textwidth]{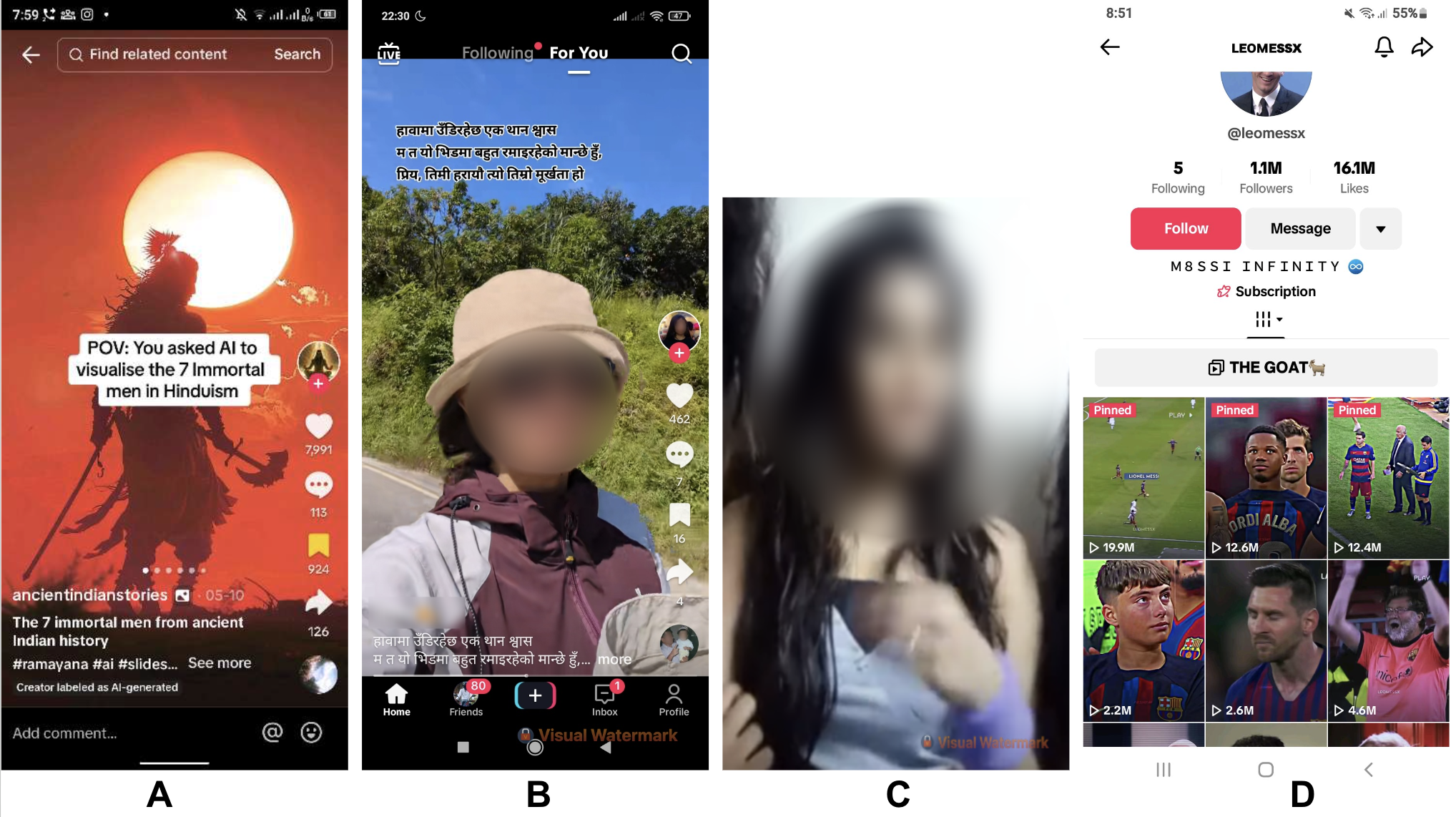} 
    \caption{ \centering Figures illustrating the diverse range of content that participants engage with on TikTok.
    \textbf{A}: Exploring AI Interpretations of Hindu Deities.
\textbf{B}: Empowered by Inspiring Quotes That Lift My Spirits. 
\textbf{C}: Reported Misleading Health Information.
\textbf{D}: Reported a Fake Messi Account
    }
    \label{fig:usersubbmittedimg}
\end{figure}


\subsection{Content Experiences on TikTok}
\label{sec:contentpref}
\changed{In addition to usage motivations, users’ experiences were deeply influenced by the types of content they encountered and engaged with on TikTok, both positively and negatively.}
When asked about their primary TikTok usage, most participants (66.67\%) used it for entertainment, followed by product discovery (13.89\%), shopping (8.33\%), and other minor uses.
Those include book recommendations, creating and watching content (including live sessions for entertainment and income), tech content, Marvel-related videos, online game streaming, life updates, and general boredom. 
\looseness -1

Overall many users enjoy TikTok.
While 31.72\% reported having favorite TikTok content,
38.88\% encountered inappropriate content. 
Yet only 4 of 42 participants shared evidence of such content, indicating a reluctance to report, possibly due to trust or moderation issues.


\looseness -1

Participants saved content that informed, inspired, or entertained them, emphasizing personal relevance and cultural ties (see Figure \ref{fig:usersubbmittedimg}).
See \changed{Supplemental Note 2} for additional user-submitted files and descriptions. Many emphasized connections to their experiences, such as one user who said, \textit{“It was related to PCOS-friendly \footnote{PCOS stands for Polycystic Ovary Syndrome, a hormonal condition that affects around 8-13\% of women in the world according to a WHO estimate.} food. I’d do anything to make the experience easier for me.”}
Travel content also resonated, with a participant saying, \textit{“I enjoyed TikTok travel videos because it was my way of seeing places around the world.”} Cultural significance emerged in responses like, \textit{“As a Hindu, I found AI versions of our gods, which amazed me.”} 

On the other hand, participants expressed alarm over various TikTok videos promoting misinformation and inappropriate behavior. 
One user described encountering perverted comments on a video where a girl danced, \textit{``People were commenting in so perverted ways. If I could, I would have reported all those men and banned from TikTok.''}
Others reported disturbing content, such as explicit pranks, animal abuse, and children engaging in inappropriate activities. They saw the need to protect others, especially young kids, from inappropriate content:

\begin{displayquote}
\textit{``It was inappropriate or concerning. The reason is we are learners, and our parents' duty is to teach us about our culture and moral values. But in this video, it was something that is quite unacceptable in our society.''}
\end{displayquote}

\begin{figure}
    \centering
    \includegraphics[width=\textwidth]{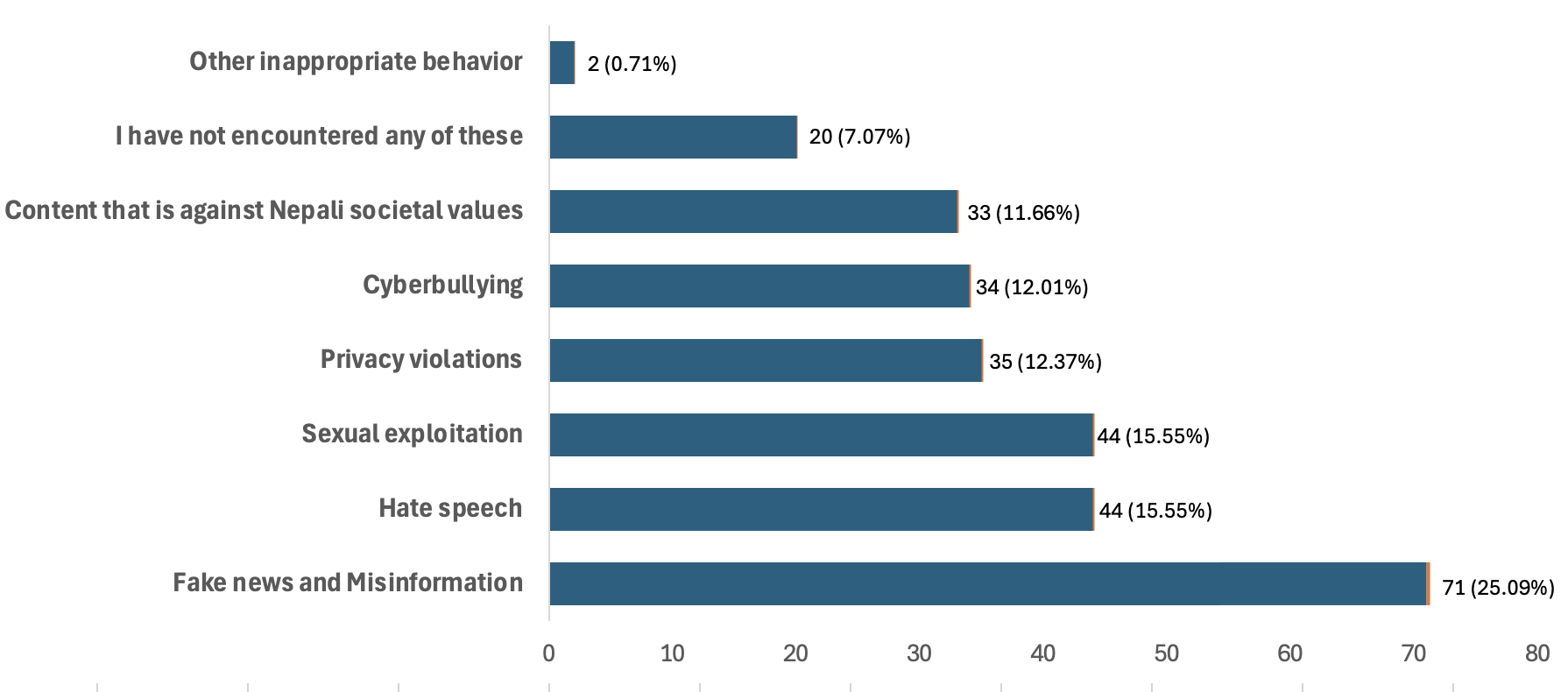}
    \caption{\finalrev{Reported Online Issues Among TikTok Users in Nepal}}
    \label{fig:tiktok_concerns}
\end{figure}

The top concern was fake news and misinformation (25.09\%), followed by hate speech and sexual exploitation (15.55\% each). Furthermore, content against Nepali societal values being flagged by 11.66\% of users suggests that cultural sensitivities are an important aspect of the TikTok experience in Nepal (\finalrev{see} Figure \ref{fig:tiktok_concerns}).

\subsection{Behavioral Shifts and Socioeconomic Patterns}
\label{sec:socialmediausage}

\changed{While users found ways to circumvent the ban and continue accessing TikTok, the restriction nonetheless prompted noticeable shifts in their broader social media habits (see Table \ref{tab:tiktok_ban}).} 
Participants reported spending more time on alternative platforms like Facebook, Instagram, and YouTube (M = 5.32, SD = 1.40) along with exposure to a wider variety of content (M = 5.37, SD = 1.23). Despite this, there was minimal reduction in overall social media usage (M = 3.99, SD = 1.68). Participants did not report a significant move toward offline activities (M = 3.96, SD = 1.57) and feelings of disconnection from communities remained low (M = 2.80).  Thus, TikTok's absence did not meaningfully impact social ties or reduce their engagement with online spaces.

\changed{However, the anticipation of the ban did prompt behavioral adjustments.} \changed{A total of} 40.7\% (44 out of 108) participants reported modifying their social media practices upon learning about the impending restriction. Some archived videos for personal use while others used the opportunity to reduce screen time or shift focus to offline activities (see Table \ref{tab:tiktok_usage_changes}). At a broader level, this reflects how quickly users recalibrate their social media habits when faced with restrictions.

\changed{Importantly, the ways users adapted were not uniform.
Although most participants reported having reliable home WiFi, this surface-level access masks deeper disparities in how different users engage with digital platforms. Participants from lower income brackets (e.g., below NPR 40,000) described using platforms for entertainment and basic connectivity with limited references to content creation, professional networking, or multi-platform use. In contrast, higher-income participants (e.g., NPR 80,000 and above) 
show more diversified, purposeful use and described more strategic and multi-functional use ranging from skill building (e.g, gardening, embroidery), freelancing, business promotion to tech news engagement. Among middle-income users, multi-platform use was superficial, driven by content format, not platform loyalty. Users switched platforms (e.g., Instagram and YouTube) mainly to replicate TikTok-style short videos, not to explore new communities. Additionally, those relying on mobile data or public internet often reported limited or single-platform usage. }

\begin{table}[t]
    \centering
    \small
    \caption{Summary Statistics of Digital Habits and Platform Use After the TikTok Ban}
    \begin{tabular}{@{}lcc@{}}
        \toprule
        \textbf{Statements: How much do you agree with the following statements? 
} & \textbf{Mean ± Std.} \\ \midrule
        1. I spend less time on social media overall since the ban & 3.99 ± 1.68 \\ 
        2. I spend more time on other platforms like Facebook, Instagram, or YouTube & 5.32 ± 1.40 \\ 
        3. I have not noticed any change in my behavior regarding social media usage & 4.55 ± 1.61 \\ 
        4. I now consume a wider variety of content from different platforms 
& 5.37 ± 1.23 \\ 
        5. I focus more on offline activities than social media & 3.96 ± 1.57 \\ 
        6. I feel disconnected from my friends and communities without TikTok & 2.80 ± 1.68 \\ 
        7. My daily routine has become more structured and productive & 4.24 ± 1.45 \\ 
        8. I found it harder to fill the time I used to spend on TikTok with other activities & 3.06 ± 1.73 \\ 
        \bottomrule
    \end{tabular}
    \label{tab:tiktok_ban}
\end{table}

\begin{table}[b]
\footnotesize
\centering
\caption{Participant-Reported Changes in TikTok Use Following Ban Announcement: Patterns of Adaptation {\scriptsize\textit{N} refers to the number of instances where participants mentioned each response pattern in their survey answers.}}
\label{tab:tiktok_usage_changes}
\begin{tabular}{@{}p{4.5cm} c p{8.5cm}@{}}
\toprule
\textbf{Response Group} & \textbf{N} & \textbf{Participant Quotes} \\ 
\midrule
Increased use of alternative platforms (Instagram, YouTube, Facebook) & 12 & \textit{“As a responsible citizen, I follow the government's decision, though it affects creative content creators. I switched from TikTok to other platforms like Facebook and Instagram Reels for entertainment.”} \\ 

Saved videos or created content before the ban & 11 & \textit{“I saved my videos from drafts, thinking they would get lost... encouraged [followers] to connect with me on other social media so we wouldn’t lose touch after the ban.”} \\ 

Reduced or stopped TikTok usage & 10 & \textit{“I just stopped using TikTok because it wasn't good for my personal growth.”} \\ 

Engaged in offline activities (workouts, spending time with friends) & 6 & \textit{“I stopped creating videos whenever I went out and felt less stressed about my viewers. Instead, I began enjoying my surroundings more, without worrying about how to capture the moment for TikTok.”} \\ 

Minimal impact or no change in usage & 5 & \textit{“The TikTok ban didn’t change much for me... I left it as is.”} \\ 

Stopped creating content due to stress or loss of interest & 4 & \textit{“I stopped making videos and stressed less about my viewers.”} \\ 

Concerns about data privacy & 2 & \textit{“I removed all the videos that were public on my profile.”} \\ 

Uninstalled or planned to uninstall the app & 3 & \textit{“I uninstalled TikTok after hearing about the ban.”} \\
\bottomrule
\end{tabular}
\label{tab:tiktok_usage_changes}
\end{table}


\subsection{
Public Perception of Nepal's TikTok Ban: Effectiveness, Justification \& Alternatives}
\label{sec:socialharmony}

\begin{figure}[b]
    \centering
    \includegraphics[width=\textwidth]{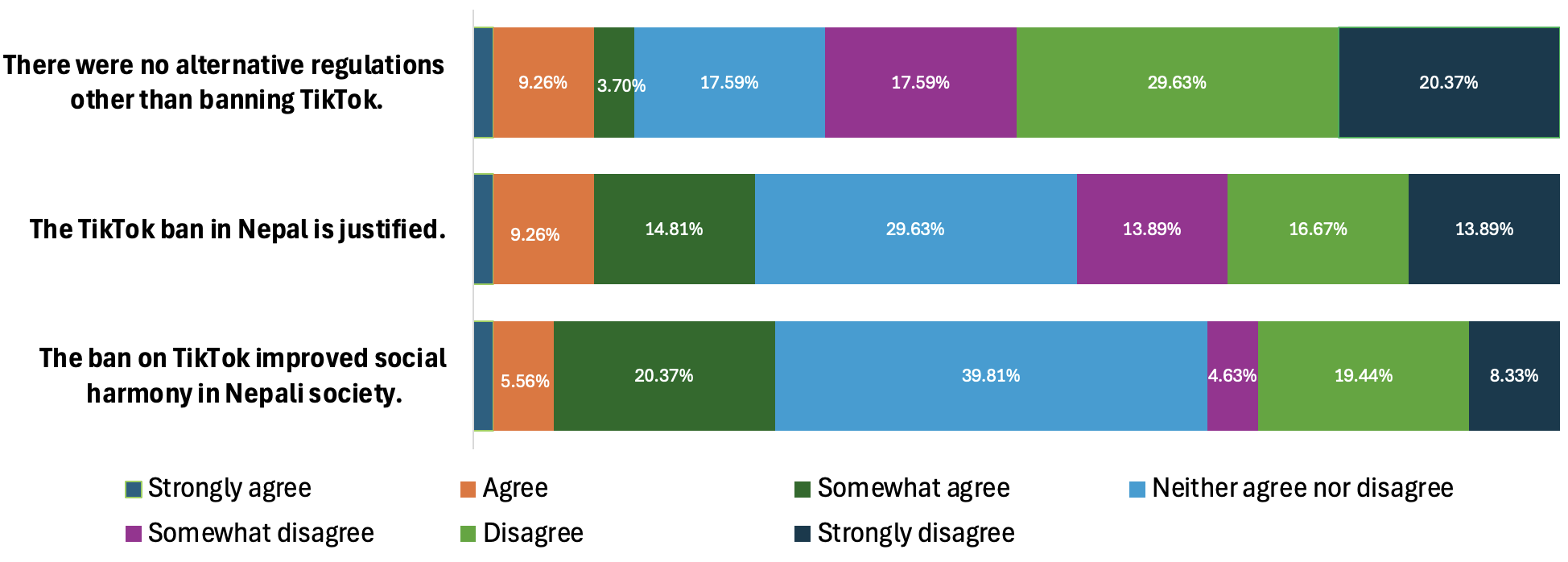}
    \caption{\centering Public Perceptions of Nepal’s TikTok Ban: \finalrev{Motives, Impacts, and Regulatory Alternatives}}
    \label{fig:harmony}
\end{figure}

\changed{While individual responses show how users adapted to the ban, it is equally important to understand how the public perceived the ban itself—its effectiveness, justification, and the potential for alternative approaches.} Overall, participants expressed skepticism about the ban.
63.89\% felt it violated freedom of expression, while many questioned whether banning was the only solution. Around 30\% disagreed with the ban's justification, and a similar share remained neutral (see Figure \ref{fig:harmony}).
Only a small 1.85\% believed that the ban improved social harmony, felt it was justified, or viewed banning as the only solution. 

\changed{
Participants attributed the TikTok ban in Nepal to a range of concerns, the most prominent being political control and suppression (47), reflecting perceptions that the ban aimed to silence dissent and limit emerging political voices. A participant shared,}
\textit{\changed{``The government banned TikTok to silence Durga Prasain’s \footnote{\changed{Durga Prasai is a controversial Nepali businessman and political activist known for his outspoken monarchist views and criticism of mainstream political parties. He gained national attention through anti-government protests and his online presence, including on TikTok \cite{civicus2024nepal}, where he shared his political opinions and mobilized public sentiment.}} voice from reaching a larger audience, fearing that more people would join the protests in support of him.''}} 
\changed{Others mentioned the platform’s alleged negative societal effects, including the spread of inappropriate or vulgar content (16), misinformation/fake news (13), and its impact on youth behavior (6). 
Cultural concerns also featured prominently, with eight (8) citing TikTok's role in eroding traditional values. As one said, \textit{“Western culture is overshadowing our values and traditions, and TikTok is fueling this shift.”} 
}
\looseness -1

\subsubsection{Prior Political Fear}
\label{sec:art}
We asked participants about experiences with censorship, including any instances where negative comments about political affiliations were monitored. Responses varied, but 
only a small fraction (7.41\%) acknowledged they felt the need to self-censor their creative work on TikTok due to fear of government repercussions.
Those who answered yes noted that strict regulations lead people to avoid sensitive political topics to avoid censorship, legal issues, or personal harm. One participant shared, \textit{``Any form of expression of creative work is always a risky thing in Nepal if it offends the people with power.''} In contrast, 37.96\% reported they often avoided political topics in their creative work.
In short, many feel free to be creative but avoid political topics due to underlying pressure.
\looseness -1

\subsubsection{User Awareness, Satisfaction, and Behavior Change with the TikTok Ban}
\label{sec:aware}

When asked about their awareness of the TikTok ban, 45.37\% were somewhat aware and 20.18\% strongly agreed. However, 14.68\% were neutral, suggesting that some lacked a full understanding of the reasons for the ban. In terms of satisfaction with the provided reasons, 33. 33\% felt neutral, while 25\% were somewhat dissatisfied and 12. 04\% were extremely dissatisfied. Only 23.15\% expressed some satisfaction, showing a general discontent with government communication. When asked whether they had been looking for other forms of entertainment or social media since the TikTok ban, a majority of respondents (39.81\%) somewhat agreed that they were looking for alternatives, while 12.96\% strongly agreed.

\subsubsection{Correlation Analysis of Awareness, Satisfaction, and Behavior}
\label{correlation}
The correlation analysis indicated a negligible and negative correlation between awareness, satisfaction, and behavior change. First, the correlation between the desire to find alternative forms of entertainment since the ban and the awareness of the reasons for the ban was minimal ($r(106) = -0.05, p = .633$). Second, the relationship between awareness of the reasons for the ban and satisfaction with those reasons was also negligible ($r(106) = -0.03, p = .757$), suggesting that being informed about the justifications did not correlate with the levels of satisfaction. Lastly, the correlation between the search for alternative entertainment and satisfaction with the ban's reasons was similarly weak ($r(106) = -0.06,\ p = .536$).

Thus, users may understand why TikTok was banned, but this awareness does not translate to major behavior changes. This shows a gap between what they know and how they feel, suggesting that simply informing users is insufficient to alleviate dissatisfaction or influence their engagement with other platforms.
\looseness -1


\subsubsection{\changed{Trust, Privacy, and Uncertainty Around Platform Re-engagement}}
\label{sec:concerns}
\changed{When asked about concerns regarding data privacy and security on TikTok}, we found a troubling persistence of concerns both before and after the ban. Prior to the ban, \changed{approximately half} (50\%), of respondents strongly expressed concern about data privacy issues; these apprehensions remained largely unchanged afterward.
We observed a statistically significant relationship between pre- and post-ban privacy concerns, $\chi^2(16) = 225.14, \; p < .001, \; \text{Cramér's} \; V = 0.72$. 
This ongoing concern indicates that the ban failed to alter users' perceptions of TikTok's safety and that users view their data privacy issues as part of broader systemic problems with TikTok, rather than it being linked to specific incidents or policy changes. 
\looseness -1

\changed{Despite these ongoing concerns, user sentiment on returning to the platform seemed mixed.}
A majority of respondents felt neutral (51\%) towards returning to TikTok, while others were somewhat (24.08\%) or very positive (17.59\%) about returning.
A small group still expressed uncertainty (3.70\%) or hesitation (2.78\%), and a
few were very negative (0.93\%) and did not want to return at all. 
\changed{These findings reflect a complex relationship between trust and platform loyalty. 
Users remain wary of TikTok’s data practices, yet many remain open to re-engagement. Together, these results indicate a need to further explore what platforms users consider as alternatives, especially when privacy concerns persist but platform dependence remains strong.}

\arrayrulecolor{black}
\begin{table}[t]
\small
    \caption{\changed{Six major themes emerged from our thematic analysis, including NA, (based on n = \textit{1407} distinct codes drawn from 648 responses across six survey questions). Note that minor overlaps exist between themes.}}
    \begin{tabular}{@{}llr@{}}
        \toprule
        \textbf{Theme}                                                            & \textbf{Definition \& Example}                                                                                                                                                                                                                                                                                                                                                         & \textbf{Code Count} \\ \midrule
        \begin{tabular}[c]{@{}l@{}}Ease-of-Use \\ \& Preference\end{tabular}      & \begin{tabular}[c]{@{}l@{}}Described an app's usability, or preferred the app over others. \\ \textit{``. . . Tiktok provides with good and more video editing} \\ \textit{options as compared to instagram and youtube.''}\end{tabular}                                                                                                                                                                  & 226            \\
        \hline
        Connections                                                               & \begin{tabular}[c]{@{}l@{}}Discussed an app's socialization, networking, trends, or virality.\\ \textit{``I use them to stay in contact with my friends and family . . .''}\end{tabular}                                                                                                                                                                                                         & 368            \\
        \hline
        Information                                                               & \begin{tabular}[c]{@{}l@{}}Commented on an app's informational content quality. \\ \textit{``I can find and read news in facebook.''}\end{tabular}                                                                                                                                                                                                                                              & 240            \\
        \hline
        Entertainment                                                             & \begin{tabular}[c]{@{}l@{}}Commented on an app's entertainment content quality.\\ \textit{``Moving from TikTok onto other platforms, some of those values} \\ \textit{feel lost. The short, fun nature of the videos on TikTok made it so}\\  \textit{one could enjoy content there with little time commitment needed.}\\  \textit{A sense of creativity existed there in ways it did not elsewhere . . .''}\end{tabular} & 205            \\
        \hline
        \begin{tabular}[c]{@{}l@{}}Diversity\\ \& Creativity\end{tabular}         & \begin{tabular}[c]{@{}l@{}}Pertained to an app's diverse reach, creative content. \\ \textit{``. . . Instagram as the age range of people using titok is really wide.} \\ \textit{My grandparent who are in their 70s and little brother who is 8,} \\ \textit{both enjoy tiktok but Instagram has very specific group of people}\\  \textit{from 15–40 (roughly) age range (sic).''}\end{tabular}                             & 237            \\
        \hline
        \begin{tabular}[c]{@{}l@{}}Privacy, \\ Security\\  \& Safety\end{tabular} & \begin{tabular}[c]{@{}l@{}}Discussed an app's privacy, data security, content report system, \\ or safety.\\ \textit{``Data security and internet safety which insurers user information} \\ \textit{safe and reliable.''}\end{tabular}                                                                                                                                                                  & 112            \\
        \hline
        NA                                                                        & \begin{tabular}[c]{@{}l@{}}Response was not valid. \\ \textit{``don't use TikTok.''}\end{tabular}                                                                                                                                                                                                                                                                                         & 99             \\ \bottomrule
    \end{tabular}

    \label{tab:thematic-analysis}
\end{table}


\label{sec:alternative}
\subsection{Alternative Platforms: Pre- and Post-Ban}
\label{altplat}

\begin{figure}[t]
    \centering
\includegraphics[width=\textwidth]{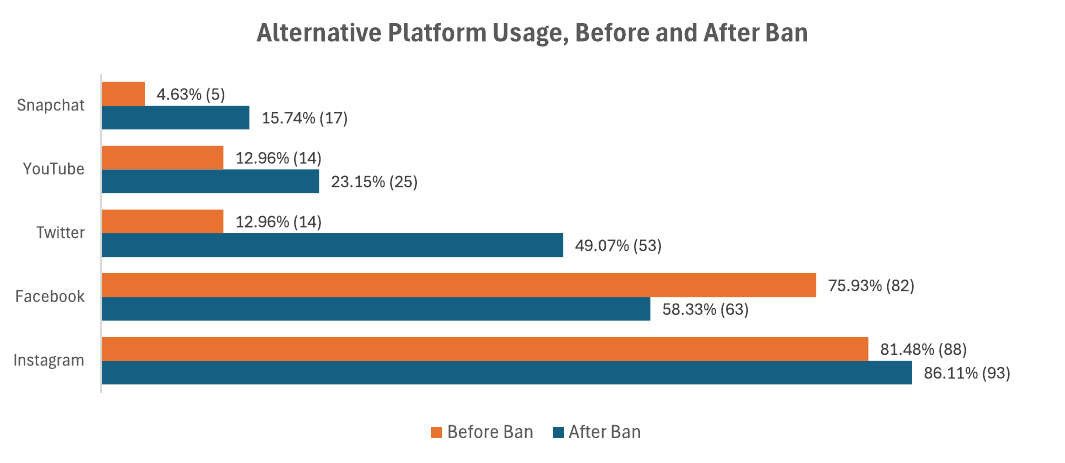}
    \caption{Most Common Alternative Platforms to TikTok (Select All That Apply)}
    \label{fig:tiktokalts}
\end{figure}

Before the TikTok ban, the top platforms were Facebook and Instagram, followed by Twitter and YouTube, with few mentions of WhatsApp, Snapchat, Dubsmash, Pinterest, and LinkedIn. Following the ban,
the most common alternatives (\finalrev{see} Figure \ref{fig:tiktokalts}) were Instagram (81\%/86\%) and Facebook (76\%/58\%). In general, many participants cited that Instagram and Facebook filled a similar niche that TikTok did due to their short video content features, both called ``Reels. An example, \textit{``Yes, they must be equivalent. YouTube, Facebook, and insta reels are almost similar to TikTok content.''}

While not nearly as popular as Instagram or Facebook, YouTube came up frequently in our free-text responses. Comparison responses frequently lump Instagram, Facebook, and YouTube together when listing similarities. In another "check all that apply" question, participants further clarified their preferences, prioritizing content variety (31.86\%) and privacy (30.88\%) as top features, followed by ease of use and user engagement (18.14\% each).

Here, we present some of the major affordances and limitations that people responded to when comparing TikTok with other alternative platforms during our thematic analysis (see Table \ref{tab:thematic-analysis}). 


\subsubsection{TikTok's Ease of Use}
\label{sec:ease}

People found TikTok easy to use, irrespective of their digital skills.
Respondents shared that this helped broaden participation in the digital space. 
For example, they shared the value of TikTok's easy-to-use video editing tools and features. 
These tools allow people to quickly create content and responses to other users (``stitches'' and ``duets''). Writing about what was lost from the TikTok ban, a respondent said, \textit{``TikTok’s duet and stitch functions enabled seamless collaboration between users, fostering creativity and interaction in a way that other platforms struggle to match.''}    
The interface was also noted as being extremely user-friendly, permitting a wider user base, \textit{``TikTok can be easily used by old-aged people.''} Another wrote about the more diverse age range TikTok catered to:
\begin{displayquote}
 \textit{``For Nepali people specifically, I think TikTok has become a comfort place than Instagram as the age range of people using TikTok is really wide. My grandparents, who are in their 70s, and my little brother, who is 8, both enjoy TikTok.''}     
\end{displayquote}

\changed{Participants also mentioned that the ease of use afforded by TikTok allowed small businesses to build an online presence without demanding much digital skills. They believe that the other alternative platforms do not give a similar preference to emerging artists and businesses like TikTok, and they mentioned that TikTok was welcoming for women creators:}
\changed{\begin{displayquote}
    \textit{``The small creators have been lost from the digital world. I think this is one major loss from the TikTok ban. Small businesses have also been impacted by the transition. TikTok was very easy to use, so small businesses used it a lot to find themselves some digital presence.''}
\end{displayquote}}



\subsubsection{\changed{TikTok's Attention-Grabbing Algorithm}}
\label{sec:tiktokalgo}

Another missed feature of TikTok during the ban was its recommendation system. Many participants, even while expressing distaste for alternative apps, mentioned its recommendation abilities that other platforms could not match---\finalrev{one shared},
\begin{displayquote}
    \textit{``Alternatives to it really are just not as great in my opinion. Sure, they might have all the functions of TikTok, like video sharing, but its algorithm is special and really works so well for the user's wants. TikTok really focuses on short, fun videos and has a great sense of community that really inspires creativity not found in most alternatives.''}
\end{displayquote}

Many respondents emphasized the algorithm's ability to make something viral, \textit{``TikTok has some unique kinds of algorithm that can make any users with less followers reach the wider audience and make the content viral in a fast and easier way''}.
\changed{Indeed, this viral aspect of TikTok was mentioned as a valuable element for growing a business, especially in a developing digital economy, where it enabled users to learn about and engage in e-commerce. Roughly 15\% of the respondents brought up the impact on the reach of local businesses while talking about the values that were lost in the transition to other platforms, 
\begin{displayquote}
    \textit{``In Nepal, mostly, people used TikTok to grow their businesses. Even where my dad used to work [at a clothing shop], they used TikTok to promote and advertise their product. They were generating a good source of income. People used to visit the store by watching videos and ads on TikTok. However, after that ban, their business was impacted a lot. People were learning to do E-commerce. E-commerce is still growing in Nepal...''}
    \end{displayquote}}

While TikTok's ability to make videos viral was seen in a positive light, some were concerned about its power, \textit{``Tiktok is addictive. For me personally I could shut down using [other apps] as soon as I wanted. Tiktok made me scroll a lot.''} Because the viral content can come from other places as well, some found it problematic, as heard in one of the responses, \textit{``Tiktok is platform where most of the contents are not appropriate for nepalese society and [I] find instagram and youtube more social friendly''}.  
This echoes one of the primary reasons for the ban stated by the government. Additional complaints viewed the algorithm as simply random, with one participant stating, \textit{``I found other platforms less chaotic and calm since I could choose what type of content I could watch.''}

\subsubsection{Other Platforms' Affordances}
While many participants found Facebook/Instagram Reels to be similar to TikTok, the long-form content on platforms such as YouTube highlighted a key affordance that TikTok lacked. Most participants found other platforms better for news and education (e.g., Twitter). As one respondent stated, \textit{``I use YouTube to watch educational videos related to my semester course.''} Further, the additional communication features (e.g., (video-) calling and messaging) available in these apps made them more valuable than TikTok. 
For example, a participant emphasized the dual value of using an alternative platform, \textit{``. . . mainly for reels and connecting to friends and family[.]
Instagram had some similar features to tiktok, like short videos which [I] felt [were] nicer.''} 
\looseness -1

Further, participants felt more assured in other platforms in terms of security and safety. One respondent highlighted TikTok's use of data, \textit{``you feel like they are somehow getting your data and using it against us''}. 
These concerns about data security and safety led individuals to place greater trust in other platforms, supporting the findings from our earlier survey quantitative responses.
We observed this in responses that contrasted TikTok with other platforms, \textit{``. . . we can also create real in Instagram just as in tiktok and it's more safe . . .''}

\section{Discussion}
\label{Discussion}
We examined the impact of the TikTok ban in Nepal through a study conducted shortly after the ban was lifted in September 2024. This allowed us to understand perceptions of the ban and people's experiences without placing them in a difficult position to talk about how they may have navigated during the ban. Our study \changed{offers a lens into how global platforms like TikTok are localized, adapted, and perceived in resource-limited contexts, where social media infrastructures are both fragile and deeply embedded in everyday life.} Our findings reveal that TikTok played a significant role in the everyday lives of Nepali users—especially the youth—serving as a space for humor, identity formation, \changed{business promotion}, and social connection (Section \ref{sec:perceivedimpacts} and Section \ref{sec:ease}). 
This echoes prior research on TikTok’s capacity to foster social expression and affective belonging in digitally precarious contexts \cite{vaterlaus2021tiktok,arias2019does,klug2022tiktok, le2021s, schaadhardt2023laughing}, and \changed{hence provides an empirical foundation to answer \textbf{RQ1}.}
\looseness -1

In the wake of the ban, digitally literate users adapted by using VPNs (Section \ref{sec:vpn}), migrating to other platforms, or pausing social media altogether (Section \ref{sec:socialmediausage}.) 
\changed{These adaptive behaviors reflect resilience and fragmentation, revealing how individuals in resource-constrained environments navigate the tension between top-down regulatory control and their desire for connection, creativity, and inclusion. Importantly, this digital adaptation is stratified: only those with sufficient technological know-how and access can bypass restrictions, exacerbating digital inequalities. Without understanding these lived experiences}, global policy discussions around social media regulation risk being skewed toward the experiences of the smaller but prominent populations in the West, overlooking the profound economic, social, and psychological impacts that such measures can have on the majority.

\changed{In the context of Nepal, these broader concerns were echoed in the public's reactions to the ban, where perceptions of government control and skepticism over official justifications came to the forefront \cite{bannepal2}.}
Participants viewed the ban in mixed ways, with many feeling it did not change their views on the app’s safety or value. 
They were largely skeptical of the government's “social harmony” rationale, interpreting it as politically driven. This distrust meant that the ban ultimately did not achieve its intended effect, directly addressing \textbf{RQ2}. There was a gap between their understanding of the ban’s reasons and their emotional reactions, as simply explaining the ban did not reduce dissatisfaction (Section \ref{sec:socialharmony}).

\changed{In response to the ban, users migrated to alternatives, mainly Instagram and Facebook, which offered similar short video features like Reels (Section \ref{altplat}). Despite these shifts, nearly a quarter of participants returned to TikTok post-ban, drawn to its unique social and creative features that other platforms could not fully replicate, directly addressing \textbf{RQ3} on post-ban social media use.}


\subsection{\changed{Migration Behaviors, and Infrastructural Precarity}}
\label{sec:discontent}
\changed{Understanding user response post-ban highlights how digital adaptation in Nepal differs from platform migration patterns observed in other global contexts. While users did not engage in overt collective protest, many practiced passive non-compliance—initially reducing engagement, then bypassing the ban through VPNs—similar to how banned Reddit communities sustained activity post-moderation \cite{horta2021platform}. Importantly, unlike the ideologically motivated and radicalized migrations observed in communities like r/The\_Donald and r/Incels, Nepali users primarily engaged with TikTok for entertainment and communication, with little evidence of politicization or toxicity.}
\changed{This distinction is crucial: whereas \citet{horta2021platform} highlights how bans can shrink but intensify extremist communities, our findings suggest that user re-engagement in Nepal was not driven by ideological commitment but by platform familiarity and cultural resonance.}

\changed{Moreover, the circumvention behavior not only reflects a shift in platform loyalty and the persistence of platform affinity, but also implies a level of technological fluency not universally accessible (Section \ref{sec:vpn}). Inadvertently, the VPN-mediated resistance may have reinforced digital inequalities in Nepal, favoring more resourced users who could afford faster internet, better devices, or exposure to global digital trends. Conversely, those reliant on mobile data or public internet faced more constrained participation, highlighting how infrastructural and economic precarity can limit meaningful digital participation.
This suggests that socioeconomic privilege shapes both access to digital platforms and the ability to fully explore, navigate, and benefit from platform affordances.}
\looseness -1

The ban raises essential questions about how societies interpret and respond to the regulation of influential technologies. If a platform as influential as TikTok can be restricted, \textit{\finalrev{W}hat does this signify for other technologies that similarly shape communication?} \changed{Our findings highlight not only the fragility of user trust in government-led digital governance but also the resilience of digital communities navigating disruptions through creative—albeit fragmented—means.}

\changed{Still, adapting to the ban was not just about finding a new app. For some respondents, TikTok was a platform to elevate local businesses and creators into an online market space, rapidly increasing their reach and economic mobility in a region where e-commerce is still growing in its early stages (Section \ref{sec:ease} and Section \ref{sec:tiktokalgo}). The sudden ban left small businesses and emerging artists with little time/resources to transition to other platforms, disrupting their visibility and growth trajectories}. 

\changed{Our findings align with prior research on platform migration behaviors. Similar to how \citet{fiesler2020moving} observed that fandom communities migrate when platforms no longer meet their needs, Nepali TikTok users adapted in creative and complex ways, reassembling their digital presence across platforms like Instagram Reels and YouTube Shorts (Section \ref{altplat}). Unlike fandom, whose migration is often gradual and community-led, the TikTok ban prompted a sudden and enforced transition, leaving Nepali users with limited time to adapt. 
This resulted in what \citet{zhang2022chinese} describe as infrastructural fragmentation---a dispersed reassembly of digital presence across multiple platforms, with no single replacement offering TikTok's reach or cohesion. Our study further supports Zhang et al.’s notion of platform precarity, as users were forced to navigate multiple spaces to retain visibility and engagement. This migration was not uniform, with many expressing uncertainty and reduced content activity, particularly among a group of creators who lacked access to alternative infrastructures.} Our findings suggest that the ban transcended a simple restriction of access; it disrupted a socio-technical ecosystem that users had come to rely on, igniting feelings of loss and frustration among its user base.

\subsection{\changed{Moving from Bans to Collective Governance Model}}
\label{sec:discontent}
\changed{
The TikTok ban in Nepal was an act of top-down regulation, imposed without public engagement or following technological or infrastructural investment. }
Despite the ban's political undertones, no efforts were made to reform tech policy, infrastructure, or engage with TikTok's parent company, ByteDance. In the absence of such systemic or corresponding social changes, users did not report a shift in trust toward the platform (Sections \ref{sec:concerns}). 
In fact, many Nepali users expressed skepticism toward the government's rationale of maintaining “social harmony”, perceiving this justification as insufficient or politically motivated. 
\changed{As our findings show, the government's ban failed to address core issues faced by Nepali users---data privacy and user safety---and did little to support those affected by the ban.}

\changed{Thus, we highlight a tension around governance involving three major actors: corporations that own and operate large platforms like TikTok, the state, and the users. As \citet{leonardi2010s} argues, the implementation of technology is deeply influenced by social interactions and power dynamics, making it essential to consider diverse perspectives when evaluating regulation and control. 
In Nepal, this tension plays out as corporate and state-led models of platform governance remain misaligned with the needs and values of the user community. 
Corporate-managed content moderation and regulation are typically centered around the corporate entity's priorities, often opaque, and rarely attuned to local norms, languages, or socio-political dynamics. We have seen this issue in many cases, such as Facebook's disastrous role in the genocide in Myanmar \cite{yue2019weaponization}. At the same time, state intervention, such as that in Nepal, is heavy-handed with limited transparency and accountability. It also does not open pathways for future actions and broader possibilities.  In contexts like Nepal, where resources are limited and the access to digital infrastructures is uneven, a state-controlled model of regulation may cause inequities and greater exclusion as our findings suggested.}
\looseness -1


\changed{To move beyond reactive bans, we advocate for a more engaged, community-driven model for platform governance. This model seeks to integrate both technological mechanisms and the users' local social realities. 
We refer to this alternative as “collective governance,” a hybrid model that integrates local values, civil society oversight, and algorithmic enforcement mechanisms.
Collective governance seeks to provide greater agency to communities through participatory capacity and support plurality in online engagements. }

\changed{To realize collective governance, we require three mutually connected layers of socio-technical arrangements.
First, platforms like Reddit and Wikipedia demonstrate how community-based norm-setting for moderation can shape user behavior and content moderation outcomes \cite{matias2019preventing, seering2019moderator, halfaker2013rise, chancellor2016thyghgapp}.
These platforms offer mechanisms that support communities to define, debate, and enforce localized norms.
In linguistically and culturally diverse contexts like Nepal, such mechanisms could allow for the recognition of regional norms, religious sensitivities, and political satire that are otherwise misread by global content moderation systems. To facilitate participatory capacity in collective governance, these norms should be public, deliberated, and contestable, as establishing community guidelines are not straightforward, as communities are heterogeneous, and even if the values were to be inscribed, values become diluted over time \cite{ghoshal2023design}.}

\changed{Second, recognizing the importance of institutional scaffolding, we propose establishing a local governance board comprising trusted community moderators, civil society members, and technical experts.
Such a board would serve as an intermediary between users, platforms, and the state, helping to interpret global content policies in ways that reflect local sociocultural realities.
Drawing from proposals in deliberative democracy (e.g., \cite{de2023critical, gutmann2004deliberative}), this board could oversee appeals processes, publish moderation rationales, and advise on content guidelines—ensuring that rule enforcement is procedurally fair, context-aware, and transparent.}

\changed{The above two mechanisms constitute the social infrastructure of collective governance. However, our findings---particularly the limited behavioral impact of Nepal’s TikTok ban---highlight the need to complement social arrangements with technical mechanisms. 
In line with CSCW scholarship on transparency and accountability, we advocate for algorithmic audits, content tracing tools, and public-facing moderation dashboards to support trust and traceability in governance \cite{jhaver2019does, sandvig2014auditing}. 
These systems can help communities evaluate norms and hold platforms accountable to algorithmic decision practices \cite{wang2020factors, binns2018s}.}

\changed{These three layers promote a participatory and polycentric oversight structure \cite{gillespie2018custodians, gorwa2019platform}.
This model seeks to redistribute power across the three actors, specifically aligning governance with the lived realities of user communities.
There are challenges in this model, including possibilities of elite capture and pragmatic concerns of scaling across multiple platforms. However, moving towards a collective governance model advocates for a more context-sensitive, accountable, and inclusive framework for platform regulation in transitional societies like Nepal.}

\subsection{\changed{Engaging with Issues of Power}}

The TikTok ban raises larger questions around political power and digital control. This brings Ackerman's concept of the socio-technical gap into focus---while technology may facilitate control, it struggles to account for the rich social dynamics that influence technology's role in society \cite{ackerman2000intellectual}.
With the government holding power to enforce such bans without transparency, we ask: \textit{What types of political actions are possible within a technological system, and to what extent should these actions be clearly defined or made visible?} This question is central for governments worldwide that seek to regulate technology while maintaining democratic principles. 

Our recommendation for deliberative democracy builds upon \finalrev{prior} CSCW \finalrev{and HCI} systems such as Consider.it \cite{kriplean2012supporting} and CommunityCrit \cite{mahyar2018communitycrit}, which aim to encourage public dialogue and deliberation. 
\changed{However, we must attend to the underlying power dynamics that shape who gets to participate in decision-making and whose voices are legitimized in these processes.}

Technology companies are mega-corporations that have control over large technological infrastructures. In contrast, countries like Nepal have limited negotiating power and fewer options for creating alternative platforms. 
The power disparity is profound. 
\changed{Moreover, when states impose regulations without community consensus or engagement, they often fail to bring about lasting changes, as per our findings.}

\changed{Community governance, as we advocate above, requires reducing this power disparity between the corporate, state, and community actors.  
This includes both macro- and  meso-level configurations of socio-technical infrastructures as discussed above, and the micro-level actions that empower local actors and individuals. 
In particular, we advocate for building structural safeguards that provide greater agency to communities and states, and for promoting digital literacy to equip individual users with the knowledge and skills to exercise control over their digital lives.}

\subsubsection{\changed{Building Structural Safeguards}}
\changed{Participants were concerned about misinformation and inappropriate content (Section \ref{sec:contentpref} and Section \ref{sec:socialharmony}).
While this may seem well-aligned with the government's rationale for the ban \cite{washingtonpost2023tiktok}, the nationwide TikTok ban raises serious concerns, especially when viewed alongside Nepal’s recently proposed Social Media Bill \cite{civicus2025social, unesco2024review}. 
The bill would grant the government sweeping powers to license platforms, enforce vague content restrictions, and penalize users under broad notions of national interest and morality.}

\changed{Together, these developments signal a broader shift toward digital authoritarianism, where state control over online spaces is expanded under the guise of safety and public order. Such regulations risk undermining free expression, suppressing creativity, and disproportionately impacting marginalized voices, including those who rely on these platforms for visibility, economic opportunity, or broader connections. In a context where digital spaces enable creative and civic expression, blanket bans and broad regulation risk silencing dissent and reinforcing exclusion---a concern voiced by several of our respondents.} 

\changed{To counter this, the focus of regulation should shift from censorship and punitive action to building structural safeguards.
These include mandating data transparency, requiring clear appeals processes for content removals, and enforcing algorithmic accountability audits \cite{sandvig2014auditing, jhaver2019does}. These approaches align with global digital rights frameworks (e.g., GDPR \cite{bentotahewa2022normative}, UN Guiding Principles on Business and Human Rights \cite{mares2011guiding}) and offer pathways for protecting users without curbing expression.}
\looseness -1

\subsubsection{\changed{Promoting Digital Literacy}}
\changed{The TikTok ban in Nepal highlighted how digital regulation without community capacity leads to uneven and ineffective outcomes.} 
For example, while some users bypassed the ban using VPNs and DNS changes, others were unfamiliar with these tools and lacked guidance on how to stay digitally connected. \changed{This disparity shows the critical need for digital literacy as an essential component of any effective and equitable governance strategy.}
Without a comprehensive plan that includes user education, blanket bans will continue to fail. 
\changed{As our study found, users deeply attach themselves to platforms like TikTok for emotional, creative, and economic reasons. When access is suddenly removed without alternatives or explanation, users experience disconnection and disempowerment.}

Digital literacy, in this context, must go beyond basic skills for platform navigation. As prior research has shown, users express concerns about algorithmic manipulation, online tracking, and the lack of diverse perspectives in personalized recommendations, highlighting the need for education on privacy, content moderation practices, and the civic implications of digital regulation \cite{khatiwada}. The goal is to equip users to recognize the socio-political forces shaping digital platforms, understand their access to these spaces, and critically assess the implications of government control over them.
\looseness -1

\changed{Designers and technologists have a key role to play here. Platforms can offer multilingual help centers, easy-to-understand explanations for moderation actions, and localized guides on secure browsing, privacy settings, and access continuity. Technologists and educators can collaborate with community organizations to develop culturally relevant curricula, peer learning opportunities, and public campaigns on digital rights and responsibilities.}

\subsection{Limitations \finalrev{and Future Work}}
This exploratory study seeks to understand Nepal's specific context, which is thus not generalizable. Nonetheless, the findings offer transferable insights that may inform broader discussions about platform bans, governance, and user behavior \changed{in other resource-constrained or transitional digital contexts}. The use of convenience sampling in the digital survey limits the representativeness of the broader population of Nepali individuals affected by the TikTok ban.
\changed{Although a power analysis confirmed the adequacy of our sample size (\textit{n = 108}), our participant pool was primarily composed of urban student populations. 
This limits the diversity of perspectives in a country marked with significant rural-urban, socio-economic, and linguistic differences. Future research should include more diverse, especially rural, voices for broader insight.
We further acknowledge potential biases, including self-selection bias and self-reporting bias in the survey responses.}

\changed{
Additionally, the stability of the results may be influenced by framing effects and respondent biases, particularly given the sensitive nature of topics like government censorship and freedom of expression. These factors could affect how participants perceive and respond to survey items, potentially impacting the consistency and generalizability of the findings if the survey were to be administered under different conditions or at a different time.} Our intent in this study was not to establish statistically significant effects or to conduct a comparative analysis across different populations. \changed{Instead, we offer a descriptive and interpretive account of how the TikTok ban was experienced and understood by a segment of Nepali users.} We also did not examine long-term behavioral shifts following the ban, \changed{which may emerge only over time.} 

Future research could build on our findings by examining such behavioral shifts through longitudinal analyses of \finalrev{cross-platform user activity before and after platform bans; changes in content consumption, migration, economic fallout for creators; as well as their digital well-being. Such work is crucial for understanding how these shifts influence not only individual livelihoods but also broader dynamics of social harmony and societal cohesion.}
\finalrev{These directions are especially important in regions like Nepal, where platform governance decisions, such as sudden bans, can disrupt digital labor ecosystems and disproportionately affect marginalized creators who rely on social media for income and visibility. It also uncovers the risks of over-reliance on a single platform, highlighting the need for more resilient strategies among creators and businesses navigating volatile digital environments.}
\finalrev{To concretely investigate these risks, especially the economic impacts on content creators, platform analytics (e.g., follower loss, engagement drop) can be combined with interviews or income self-reports to understand how bans affect creators who rely on social media for their livelihood.}
\looseness -1


\finalrev{Moreover,} understanding user migration patterns to platforms like Instagram or Facebook after the TikTok ban can help triangulate the findings. 
\changed{Our study is also limited in that we do not engage much with socio-economic factors in examining the impact of the ban.} 
In nearby countries like India and Bangladesh, TikTok often appealed to rural and poorer populations, whereas Instagram was often associated with urban, affluent, and more curated lifestyles 
\cite{latimes2020tiktokban, tbsnews2023tiktok}. 
As we argue that the ban in Nepal did not align with the population's values and needs, we acknowledge that the ``population'' is not homogeneous. \changed{The ban can likely be felt differently by people from different socio-economic backgrounds.}
\finalrev{As such, future work could explore how socio-economic status influences platform preference and access, revealing the extended social and economic consequences of platform governance across the Global South.
}

\section{Conclusion}
Our study looks at how people in Nepal, a low-resource transitional country, perceive and react to a ban on TikTok imposed by the government. Participants shared that they switched to alternative social media platforms, demonstrating their ability to adapt and find ways to connect digitally despite facing challenges—some leverage VPNs to bypass regulatory restrictions. Similarly, considering the variation in TikTok usage, we emphasize the need to capture real perceptions about the platform and its ban from a wide range of users. Our insights may be transferable to other contexts, such as deliberating platform bans or regulatory restrictions. Our work adds to the growing call for diversifying design research beyond the West, particularly in centering the voices of those not historically considered primary users of social media platforms like TikTok. At the same time, we draw attention to power issues, particularly in contemplating how communities with limited control over globalized technological infrastructure can govern such platforms.
\looseness -1

\section{Acknowledgment}
We would like to sincerely thank Prajwal Khatiwada for managing the international compensation process and ensuring smooth coordination throughout the study. We also extend our gratitude to Prakash Bhattarai from the Centre for Social Change for his valuable time in assisting us with the cultural sensitivity analysis of the survey and guiding us throughout. We also extend our thanks to Nuva Rai of Body and Data for helping us understand the context and providing data to guide our survey. Thanks to Surochit Pokhrel for his help with advertising the study locally in Nepal. Additionally, we thank the Sensify Lab members for their valuable feedback, support during the review of this manuscript, and coordination and contributions during the pilot testing of the survey.

\bibliographystyle{ACM-Reference-Format}
\bibliography{reference, new_reference}



\end{document}